\documentclass[epj]{svjour}
\usepackage{graphicx}
\begin{document}
\title{Skyrmion Dynamics and Transverse Mobility: Skyrmion Hall Angle Reversal on 2D Periodic Substrates with dc and Biharmonic ac Drives}
\titlerunning{Skyrmion Dynamics and Transverse Mobility}
\author{N.P. Vizarim\inst{1,2} \and C.J. O. Reichhardt\inst{1} \and
P.A. Venegas\inst{3} \and C. Reichhardt\inst{1}
}                     
%
%
\institute{Theoretical Division and Center for Nonlinear Studies,
  Los Alamos National Laboratory, Los Alamos, New Mexico 87545, USA
  \and
  POSMAT - Programa de P\'{o}s-Gradua\c{c}\~{a}ao em Ci\^{e}ncia e Tecnologia
  de Materiais, Faculdade de Ci\^{e}ncias, Universidade Estadual Paulista - UNESP,
  Bauru, SP, CP 473, 17033-360, Brazil
  \and
Departamento de F\'{i}sica, Faculdade de Ci\^{e}ncias, Universidade Estadual Paulista - UNESP, Bauru, SP, CP 473, 17033-360, Brazil}
\date{Received: date / Revised version: date}
%
\abstract{
  We numerically examine the dynamics of a skyrmion interacting with a
  two-dimensional
  periodic substrate under dc and biharmonic ac drives.
We show that the Magnus force of the skyrmion 
produces circular orbits that can resonate with
the ac drive and the periodicity of the substrate
to create quantized motion both parallel and perpendicular to the
dc drive.
The skyrmion Hall angle
exhibits a series of increasing and/or decreasing steps
along with strongly fluctuating regimes.
In the phase locked regimes,
the skyrmion Hall angle is constant
and the skyrmion 
motion consists of
periodic orbits encircling
an integer number of obstacles per every or every other ac drive cycle.
We also observe phases in which
the skyrmion moves
at $90^\circ$ with respect to the driving direction
even in the presence of damping,
a phenomenon called absolute transverse mobility
that can exhibit reentrance as a function of
dc drive.
When the biharmonic ac drives
have different amplitudes, in the two directions we
find regimes in which the skyrmion Hall angle
shows a sign reversal from positive to negative,
as well as a reentrant pinning
effect in which the skyrmion is mobile at low drives but becomes pinned
at higher drives.
These behaviors  
arise due to the combination of the Magnus force
with the periodic motion of the skyrmions,
which produce Shapiro steps, directional locking, and ratchet effects.  
} 
\maketitle
\section{Introduction}
There are a number of systems that can be modeled effectively
as a particle
moving over a two-dimensional (2D) periodic substrate under a dc drive.
Such systems include  vortices in type-II superconductors
interacting with 2D pinning arrays
\cite{Harada96,Reichhardt97,Martin99,Reichhardt08a,Gutierrez09,Sadovskyy17,Ge18},
colloids moving over optical trap arrays
\cite{Korda02,MacDonald03,Bohlein12,Vanossi12,Hasnain13} 
or periodically patterned surfaces \cite{McDermott13a,Tierno09,Loehr18,Cao19,Stoop20}, 
and models of atomic friction \cite{Tekic05,Vanossi13}. 
Typically, when an additional ac drive is applied to these systems,
the particle can exhibit 
a resonance phenomena which results in steps in the 
velocity versus dc force curves.
These steps are known as Shapiro steps when the ac drive is parallel
to the dc drive
\cite{Shapiro63,Barone82,Martinoli75,vanLook99,Reichhardt00b,Dobrovolskiy15,Juniper15,Brazda17}.
On each step,
the particle velocity remains fixed even though the dc drive is changing,
so that the particle remains in resonance.  
If the ac drive is perpendicular to the dc drive,
then for 2D periodic substrates it is possible to have another type of 
phase locking distinct from Shapiro steps which is known 
as transverse phase locking \cite{Reichhardt01,Marconi03}. 
Multiple ac drives can also be applied
in the form of multiple frequencies in the same direction or
the same frequencies
in different directions and out of
phase by $90^\circ$ to create a circular drive
\cite{Reichhardt02b,Guantes03,Reichhardt03,Speer09,Chacon10,Mukhopadhyay18}.

Under a dc drive,
a particle experiencing a circular ac drive moving over
a 2D periodic substrate can not only exhibit 
phase locking effects in the
direction of drive,
but in certain cases can
move transverse
to the dc drive,
leading to a finite Hall angle due to the chiral nature of the motion \cite{Reichhardt02a}.
If the crossed ac driving
is more complex,
so that in the absence of a dc drive the particle would follow a Lissajous pattern,
the system can  exhibit absolute transverse mobility,
where the particle moves perpendicular to an applied dc drive,
or even negative mobility, where the
particle moves against the applied dc drive
\cite{Reichhardt03}.
Such phenomena have been observed for 
circular colloidal motion on magnetic bubble arrays,
where various types of localized translating quantized motion
occur \cite{Tierno07,Soba08}. 
Studies of this motion performed to date have focused on overdamped systems;
however, in other systems,
additional nondisspative effects arise, such as
a gyroscopic or Magnus force which creates velocity components perpendicular
to the net force on the particle.
Such effects can appear in superconducting vortices 
\cite{Ao93},
vortices in superfluids and Bose-Einstein condensates
\cite{Yabu97,Groszek18},
magnetic vortices \cite{Pribiag07,Bolte08},
charged particles in magnetic fields \cite{Wiersig01,Khoury08},
and active spinner systems \cite{vanZuiden16,Han17,Reichhardt19a,Yazdi02}.
Another system
where a strong Magnus force is present
is skyrmions in chiral magnets.
Skyrmions are particle-like magnetic textures
that can be set into motion
with an applied current and
that can interact with tailored pinning structures or landscapes 
\cite{Muhlbauer09,Yu10,Nagaosa13,Schulz12,Iwasaki13,Woo16,Tekic19,Xiong19}.
The Magnus 
force can strongly affect how the skyrmions move under an
external drive and in the presence of disorder or
a confining potential.
It can produce a drive-dependent skyrmion Hall angle
due to velocity-dependent asymmetric scattering of the skyrmions by defects
\cite{Reichhardt15a,Reichhardt15,Jiang17,Litzius17,Legrand17,Zeissler20},
spiraling skyrmion motion around defects
\cite{Reichhardt15a,Liu13,Muller15,Buttner15,Martinez16,GonzalezGomez19,Salimath19},
and speed up effects
\cite{Reichhardt15a,Muller17,Salimath19,CastellQueralt19,Tomasello18} where
the pinning force in combination with
the Magnus effect can accelerate the skyrmion.
Since skyrmions also show promise for various 
applications \cite{Fert17,Tomasello14,Prychunenko18},
understanding how to control skyrmion
motion in the presence of nanostructured pinning arrays could be
a promising approach for creating skyrmion based devices.   

Skyrmions driven over a one-dimensional (1D) periodic substrate
under a combined
dc and ac drive exhibit a number of phase locking phenomena
including Shapiro steps,
which occur when the ac and dc drives are parallel;
however, due to the Magnus force, skyrmions can also show Shapiro steps
in the velocity transverse to the drive \cite{Reichhardt15b}.
In an overdamped system with a 1D periodic substrate,
an ac drive applied
perpendicular to the dc drive produces no phase locking;
however, for skyrmions, the Magnus force can create
a Magnus-induced phase locking effect
in this drive configuration \cite{Reichhardt15b,Reichhardt17}.
If the 1D substrate is asymmetric,
it is possible to observe a Magnus-induced phase locked ratchet
effect which is absent in overdamped systems \cite{Reichhardt15a}.
For skyrmions 
interacting with a 2D periodic array of scattering sites,
various types of directional locking effects can occur under strictly dc
driving
\cite{Reichhardt15a,Feilhauer19,Vizarim20}.
In this
case, since the skyrmion Hall angle increases with
increasing skyrmion velocity, the changing direction of flow of the skyrmion
becomes locked to 
certain symmetry directions of the periodic substrate,
leading to a quantized skyrmion Hall angle.
There are various  methods for
producing traps or obstacles for skyrmions \cite{Stosic17,Fernandes18,Toscano19}, 
including techniques for creating periodic obstacle arrays \cite{Saha19}.
Coupling of a skyrmion 
to a periodic substrate can also be achieved
by causing the skyrmion to interact with
a lattice of superconducting vortices \cite{Menezes19,Palermo20}.
An ac drive can be applied with a current or with oscillating 
magnetic fields \cite{Chen19,Chen20}.

\begin{figure}
\includegraphics[width=3.5in]{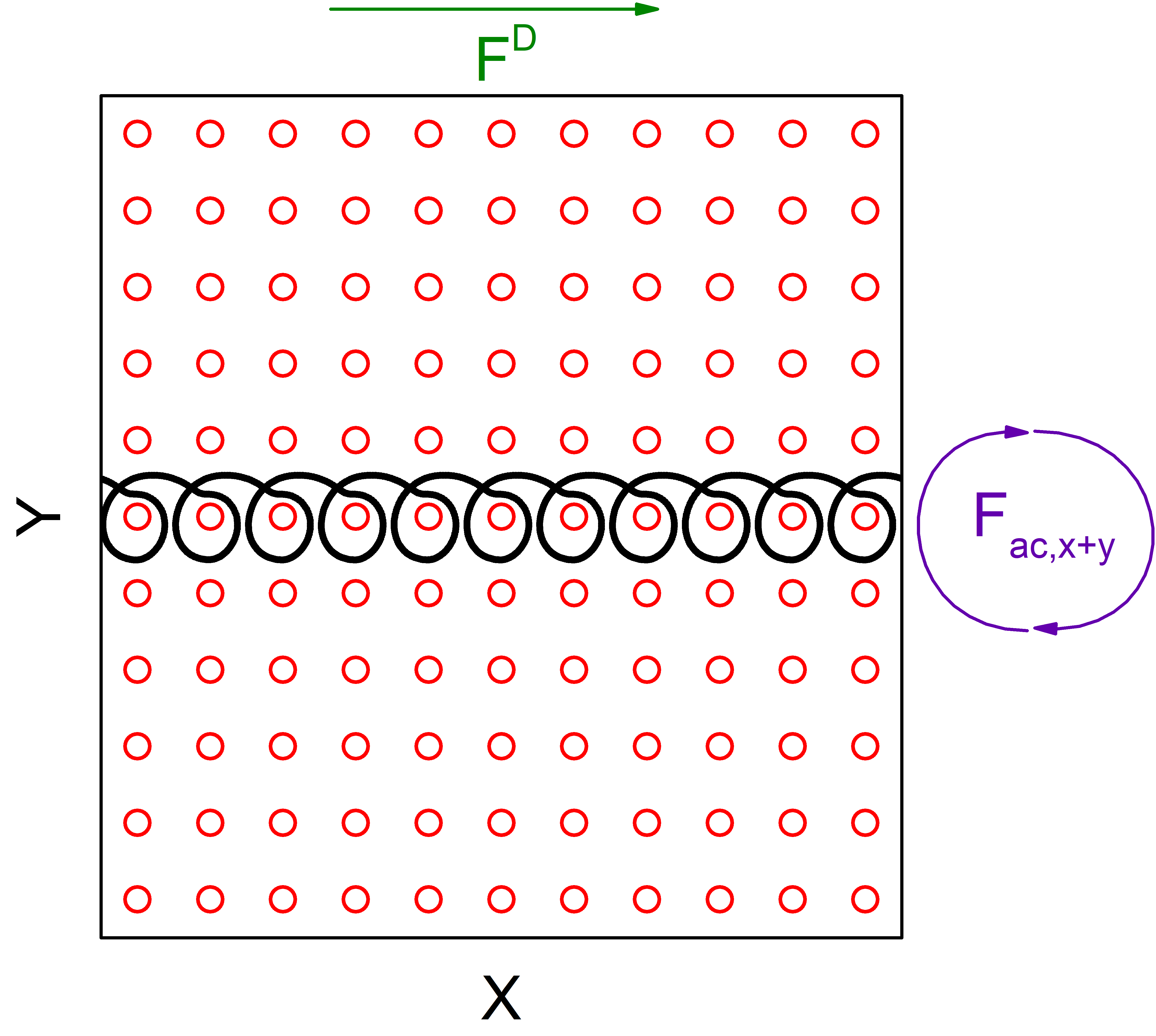}
  \caption{Image of the simulated system consisting of a square array
    of obstacles (red circles) modeled as repulsive Gaussian scattering sites.
    A skyrmion (trajectory indicated by a black dot) is subjected to
    a dc drive $F^D$ applied along the $x$ direction as well as an ac drive
    $F^{AC}=A\sin{(\omega_1 t)}{\bf \hat x} + B\cos{(\omega_2 t)}{\bf \hat y}$.
    When $\omega_1=\omega_2$, as illustrated here, the ac drive is circular.
}
\label{fig:1}
\end{figure}

In previous work, we examined skyrmions interacting with a 2D obstacle array
under only  dc driving, where we found
a series of directional locking effects that depend upon
the size of the obstacles or
whether the pinning sites are attractive or repulsive \cite{Vizarim20}.
Here we consider the same system but add a biharmonic ac drive
given by $A\sin(\omega_1 t){\bf \hat x} + B\cos(\omega_2 t){\bf \hat y}$
such that the skyrmion executes
a circular orbit as illustrated in  Fig.~\ref{fig:1}.
For varied parameters, we find that this system exhibits 
a rich variety of phase locking phenomena,
including phases where the
skyrmion motion remains locked in a particular direction
while the skyrmion orbit
encircles an integer number of obstacles
during each ac drive cycle. 
Additionally,
the skyrmion Hall angle exhibits a series of increasing and/or decreasing
steps
as a function of increasing dc drive.
We observe several phases in which the skyrmion moves at
$90^\circ$ to the dc drive
in spite of the fact that the intrinsic skyrmion Hall angle is 
much smaller than $90^\circ$,
giving an example of
transverse mobility,
and in some cases we find
a reversal of the sign of the skyrmion Hall angle from positive to negative.          

\section{Simulation}

We model a two-dimensional system  
of size $L\times L$
with periodic boundary conditions in the $x$ and $y$
directions containing a square 
array of obstacles with lattice constant $a$. 
We place a single skyrmion in the system and apply
both a dc drive and a biharmonic ac drive which creates a circular motion
of the skyrmion, as illustrated
in Fig.~\ref{fig:1}.
The equation of motion of the skyrmion 
is based upon a particle model for skyrmions
used previously to model skyrmions interacting with
pinning \cite{Vizarim20,Lin13,Brown19},
with the form
\begin{equation}
\alpha_d {\bf v}_{i} + \alpha_m {\hat z} \times {\bf v}_{i} =  {\bf F}^{obs} + {\bf F}^{DC} + {\bf F}^{AC} 
\end{equation}
Here $\alpha_{d}$ is the damping term which aligns the velocities
of the skyrmions with the net applied forces, while
${\alpha }_{m}$ is the Magnus term 
which produces skyrmion velocities
that are perpendicular to the net forces experienced by the skyrmion.
The term ${\bf F}^{obs}$ on the right represents
the interaction force between skyrmions and obstacles,  
${\bf F}^{obs}=\sum_i^{N_o}{\bf {F}}^o_i=-\mathrm{\nabla }U_o=-F_or_{io}e^{-{\left({r_{io}}/{a_o}\right)}^2}{\widehat{\bf r}}_{io}\ $,
where $F_o=2U_o/a^2_o$.
The obstacle potential energy is
$U_o=C_oe^{-{\left({r_{io}}/{a_o}\right)}^2}$, the potential strength is $C_o$, 
the distance between skyrmion $i$ and obstacle $o$ is $r_{io}$,
and the obstacle radius is $a_o$.
We cut off the skyrmion obstacle interaction at $r_{io}=2.0$,
since beyond this length the interaction is negligible. 
The obstacle density is fixed at $0.093/{\xi }^2$
and the obstacle radius is fixed at $a_0=0.65$.
The term ${\bf F}^{DC}$
represents the dc driving force 
applied along the $x$ direction, as indicated in Fig.~\ref{fig:1}. 
We increase the dc drive in small steps of $\delta F^D=0.001$,
and wait ${10}^5$ simulation 
time steps between drive increments to ensure
that the system has reached a steady state
before obtaining average velocities.
We normalize 
the damping and Magnus coefficients to ${\alpha }^2_d+{\alpha }^2_m=1$.

The ac drive has the form
\begin{equation}
{\bf F}^{AC} = A\sin(\omega_{1} t){\bf x} + B\cos(\omega_{2} t){\bf y}. 
\end{equation}
Here, $A$ and $B$ are the ac drive amplitudes and $\omega_{1,2}$ are the
ac drive frequencies.
In the first part of the work we fix
$A = B = 0.5$ and
$\omega_{1} = \omega_{2}$,
and throughout the work we fix
$\omega_1 = 2\times 10^{-4}$ in inverse simulation steps.
We measure the  skyrmion velocity parallel, $\left\langle V_{\parallel }\right\rangle $, and perpendicular, $\left\langle V_{\bot }\right\rangle $, to the dc drive.
In the absence of obstacles,
strictly dc driving combined with
the  Magnus force  causes the skyrmion to move at 
an intrinsic Hall angle of $\theta^{int}_{sk} = \arctan(\alpha_{m}/\alpha_{d})$.
When pinning
or obstacles are present, the skyrmion Hall angle
develops a drive dependence that can be  measured 
using the  ratio of the velocity transverse to the drive to the velocity parallel
to the drive,
$\theta_{sk}={\mathrm{arctan} \left(\left\langle V_{\bot }\right\rangle /\left\langle V_{\parallel }\right\rangle \right)\ }$.

\section{dc and Biharmonic ac Drives}

\begin{figure}
\includegraphics[width=3.5in]{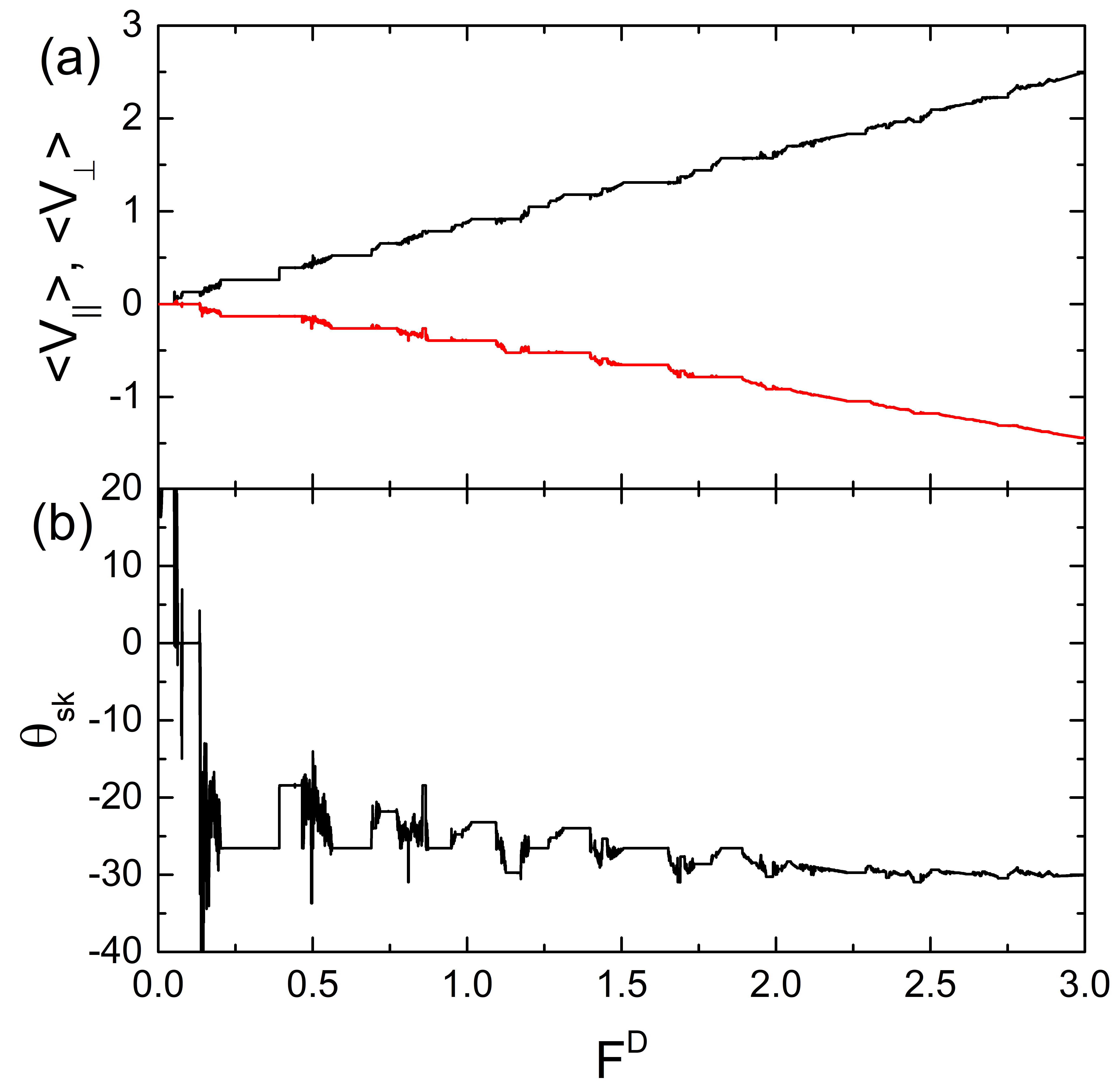}
\caption{  
  (a) $\langle V_{\perp}\rangle$ (red) and $\langle V_{||}\rangle$ (black) vs $F^{D}$
  for a system
  with $\omega_1=\omega_2$,
  $A=B=0.5$, and $\alpha_m/\alpha_d=0.577$.
(b) The corresponding $\theta_{sk}$ vs $F^{D}$.}
\label{fig:2}
\end{figure}

\begin{figure}
\includegraphics[width=3.5in]{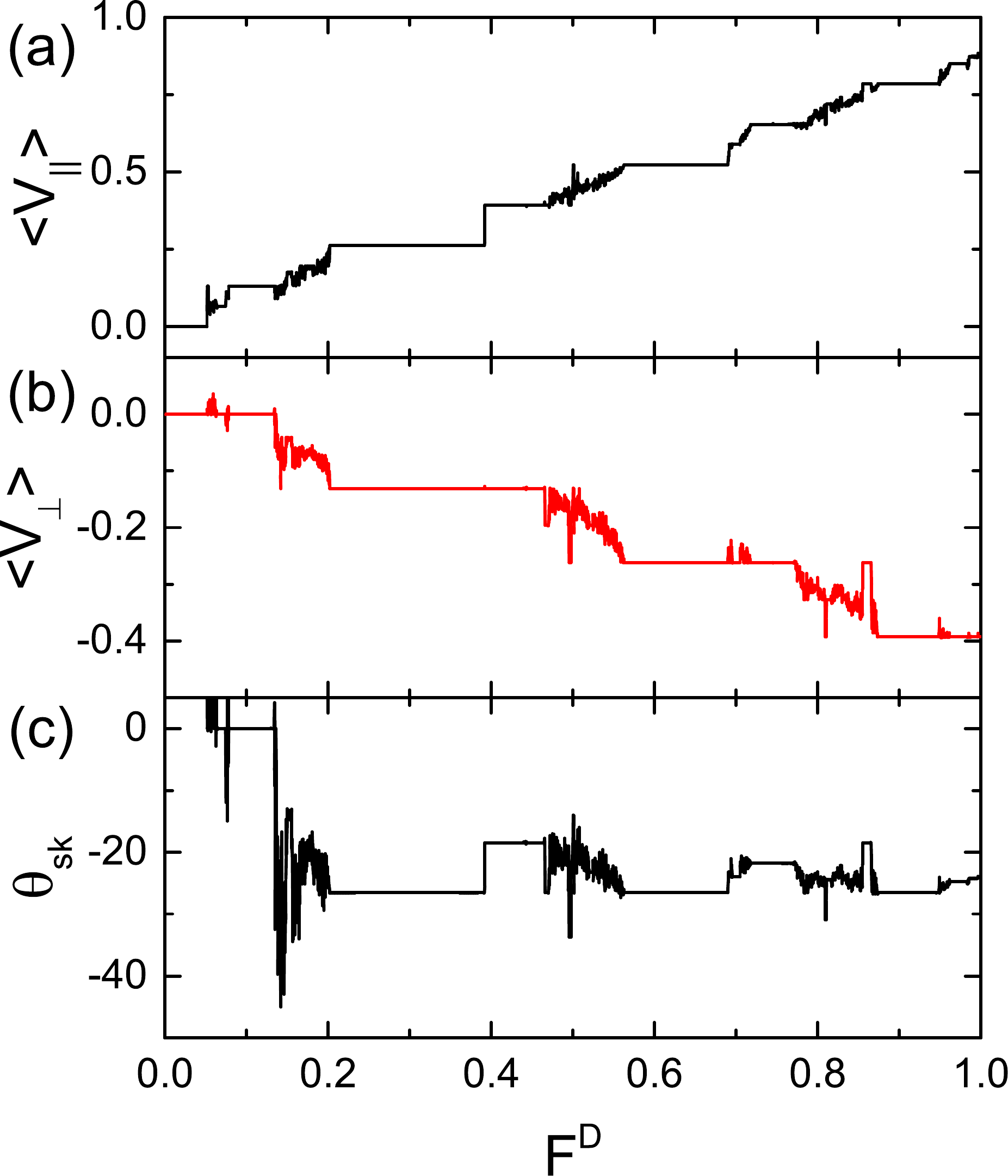}
\caption{ 
(a) $\langle V_{||}\rangle$, (b) $\langle V_{\perp}\rangle$, and  (c)
  $\theta_{sk}$ vs $F^{D}$ for the system in Fig.~\ref{fig:2}
  with $\omega_1=\omega_2$,
  $A=B=0.5$, and $\alpha_m/\alpha_d=0.577$,
 showing 
locking at $\theta_{sk} = -26.565^{\circ}$ and $\theta_{sk} = -18.43^\circ$. } 
\label{fig:3}
\end{figure}

\begin{figure}
\includegraphics[width=3.5in]{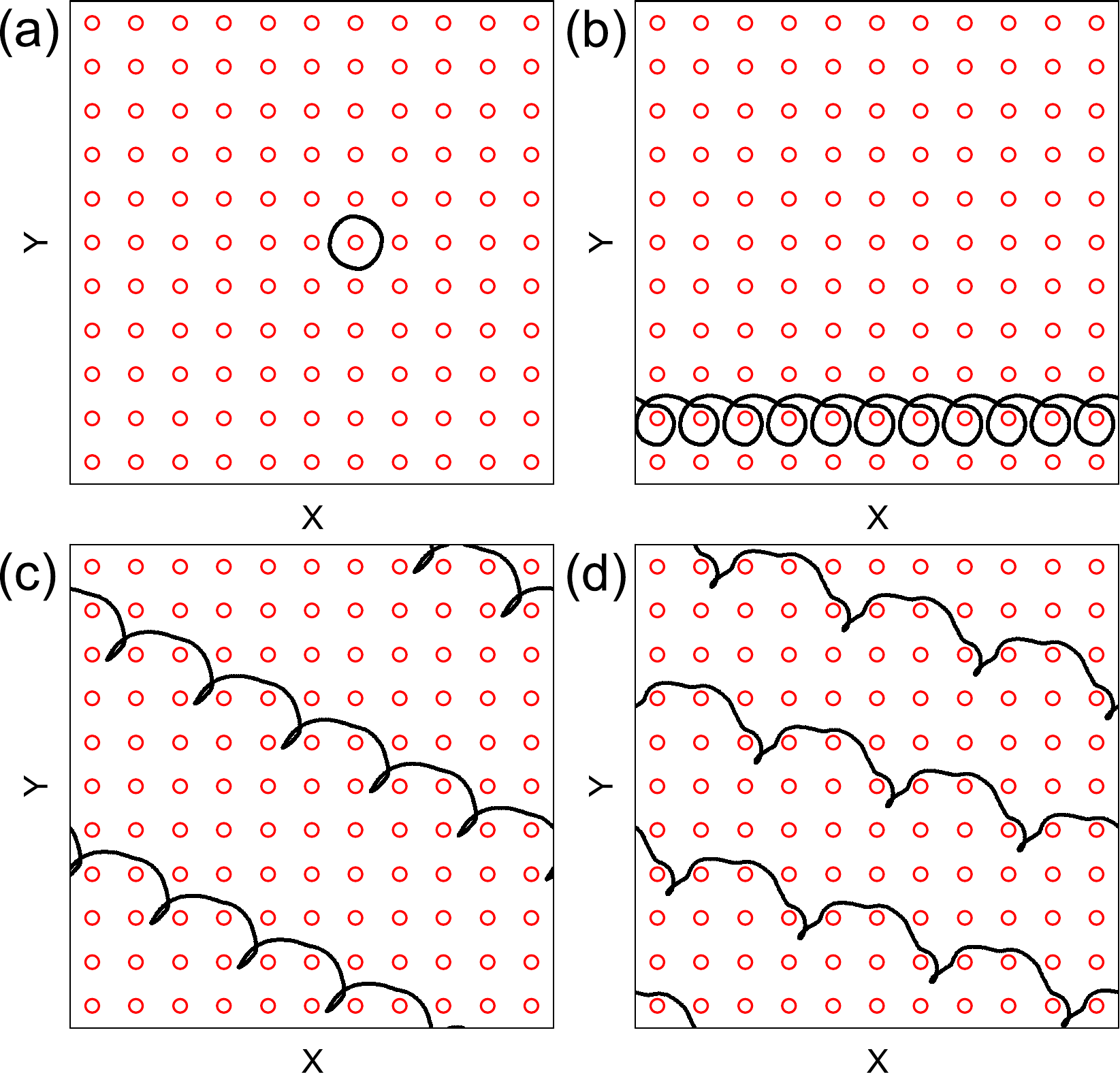}
\caption{ The obstacle locations (open circles) and skyrmion trajectory (line)
  for the system in Fig.~\ref{fig:3}
  with 
  with $\omega_1=\omega_2$,
  $A=B=0.5$, and $\alpha_m/\alpha_d=0.577$.
(a) $F^{D} = 0.028$, in the pinned phase. 
(b) $F^{D} = 0.1$, where the skyrmion motion is locked to the $x$ direction. 
  (c) $F^{D} = 0.24$, where locking occurs at $\theta_{sk} = -26.565^\circ$.
  (d) $F^{D} = 0.425$, where locking occurs at 
a reduced angle of $\theta_{sk} = -18.43^\circ$.}
\label{fig:4}
\end{figure}

In Fig.~\ref{fig:2}(a) we plot
$\langle V_{||}\rangle$ and $\langle V_{\perp}\rangle$ versus
$F^{D}$
for a system with $\alpha_{m}/\alpha_{d} = 0.577$
and $A = B = 0.5$, while in
Fig.~\ref{fig:2}(b) we show 
the corresponding $\theta_{sk}$ versus $F^{D}$.
Both $\langle V_{||}\rangle$ and $\langle V_{\perp}\rangle$
increase in a series of steps.
There is also a series of steps
in $\theta_{sk}$, 
but these steps show oscillatory jumps
both up and down, indicating
that the skyrmion Hall angle can both increase and decrease
as a function of increasing $F^{D}$.  This is in contrast to the
behavior in the absence of an ac drive
where the skyrmion Hall angle monotonically increases with increasing $F^{D}$.
The intrinsic Hall angle is $\theta^{int}_{sk} = -29.98^\circ$,
and the measured $\theta_{sk}$ gradually
approaches this intrinsic value at high $F^{D}$.
In Fig.~\ref{fig:3}(a,b,c) we show a close up of 
$\langle V_{||}\rangle$, $\langle V_{\perp}\rangle$,
and $\theta_{sk}$ versus $F_{D}$
over the range $0 < F^D < 1.0$ for the system in Fig.~\ref{fig:2}.
There is an initial pinned phase for $F^D\leq 0.075$
where $\langle V_{||}\rangle = \langle V_{\perp}\rangle = 0.0$.
In Fig.~\ref{fig:4}(a) we illustrate the
skyrmion trajectory in the pinned phase at $F^{D} = 0.028$,
where the skyrmion moves in a circular orbit around a single obstacle. 
When  $A = B = 0.0$, the ac driving is absent and
there is no pinned phase since the range of the obstacles is finite and the
skyrmion can always move between the obstacles.
Under a finite
circular ac drive,
the effective dynamical radius of the skyrmion increases,
causing the skyrmion to interact
with a larger number of obstacles during each ac drive cycle and permitting it
to become trapped even under a finite dc drive.   

In Fig.~\ref{fig:3},
$\langle V_{||}\rangle$ is finite and $\langle V_{\perp}\rangle=0$ over the range
$0.075 < F^{D} < 0.15$,
giving a skyrmion Hall angle of $\theta_{sk}=0^\circ$.
In Fig.~\ref{fig:4}(b)
we plot the skyrmion trajectory at $F^{D} = 0.1$,
where the skyrmion translates
along the $x$-direction
by one obstacle per ac cycle.
Within the range of drives for which
the velocity is locked in the $x$-direction,
it is possible to have steps in $\langle V_{||}\rangle$ on which the 
orbits are similar to those shown in Fig.~\ref{fig:4}(b)
but where the skyrmion encircles each obstacle twice in a single
ac cycle before translating
by one lattice constant in the $x$ direction.
For $0.15 \leq F^{D} < 0.2$,
the skyrmion begins to move in the $y$ direction as well and
the dynamics is more
chaotic,
with no drive interval
over which the motion is locked to a specific  
direction.
For $0.2 \leq F^{D} < 0.4$, the skyrmion motion 
is periodic and locked to an angle of 
$\theta_{sk} = -26.565^\circ$.
Here,
during each ac drive cycle, the skyrmion
translates by two lattice constants
in the $x$ direction and one in the $y$
direction, giving $\theta_{sk} = \arctan(1/2) = -26.565^\circ$.
In Fig.~\ref{fig:4}(c) we illustrate a
skyrmion trajectory in this regime at $F^{D} = 0.24$.
When $0.4 \leq F^{D} < 0.475$, the magnitude of the skyrmion Hall angle
decreases and the locking angle is
$\theta_{sk}=-18.43^\circ$,
with the skyrmion moving $3a$ in
the $x$ direction and $a$ in the $y$ direction during each ac
drive cycle, as
shown in Fig.~\ref{fig:4}(d) at $F_{D} = 0.425$.
As $F^D$ increases further,
$\theta_{sk}$ jumps between the
two main locking directions
of $\theta_{sk}=-26.565^\circ$ and $\theta_{sk}=-18.43^\circ$,
and additional fractional locking steps appear
in the velocities and the skyrmion Hall angle
corresponding to $R = 3/7$ and $R=3/8$. There are also
several regions of chaotic motion.

\begin{figure}
\includegraphics[width=3.5in]{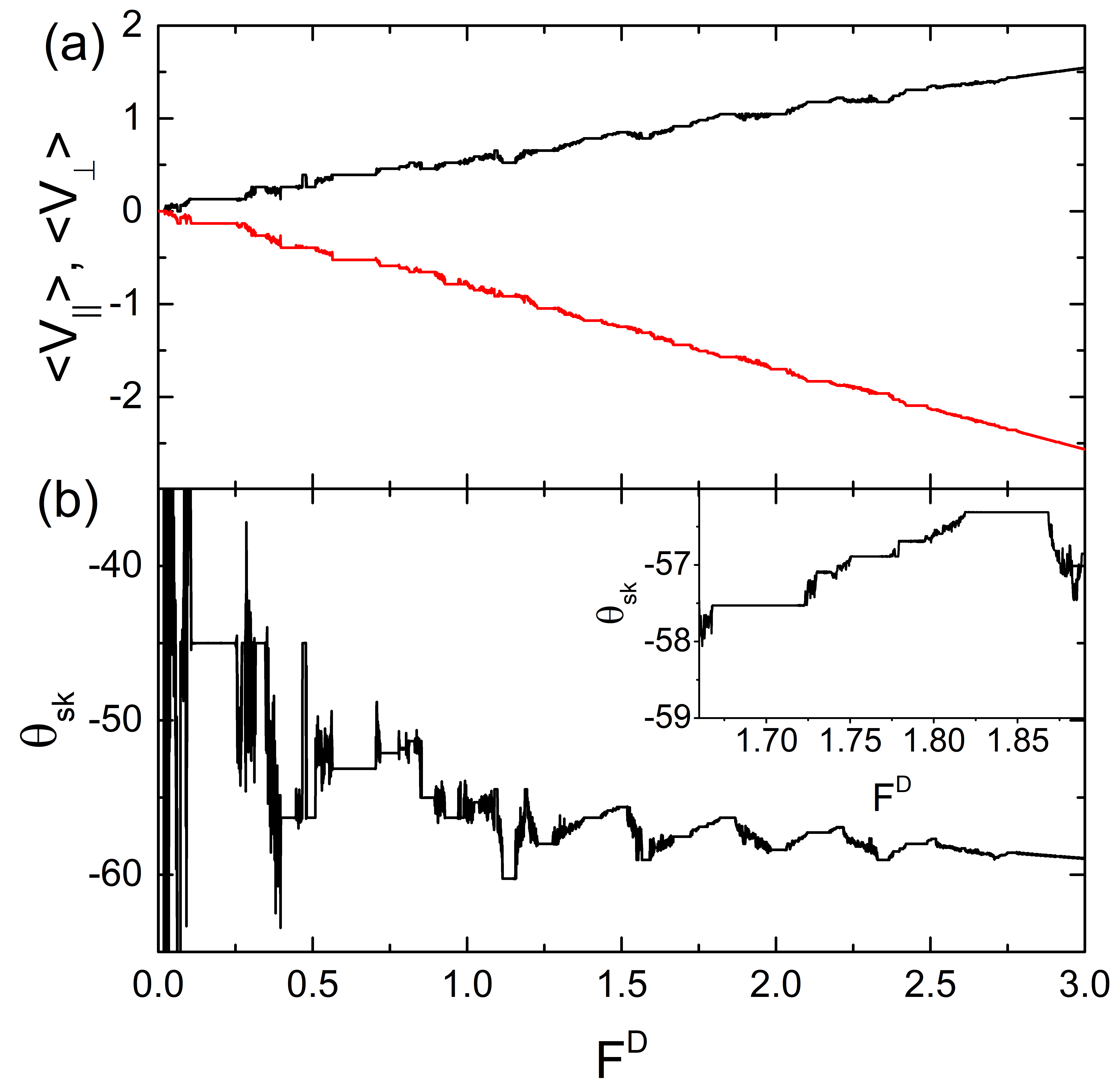}
\caption{
  (a) $\langle V_{\perp}\rangle$ (red) and $\langle V_{||}\rangle$ (black)
  vs $F_{D}$ and (b) the corresponding $\theta_{sk}$ vs $F_{D}$ for a system
  with $\alpha_{m}/\alpha_{d} = 1.732$, $\omega_1=\omega_2$,
  and
  $A = B = 0.5$. The inset of (b) shows a zoomed in view of $\theta_{sk}$ vs $F^{D}$.}   
\label{fig:5}
\end{figure}

\begin{figure}
\includegraphics[width=3.5in]{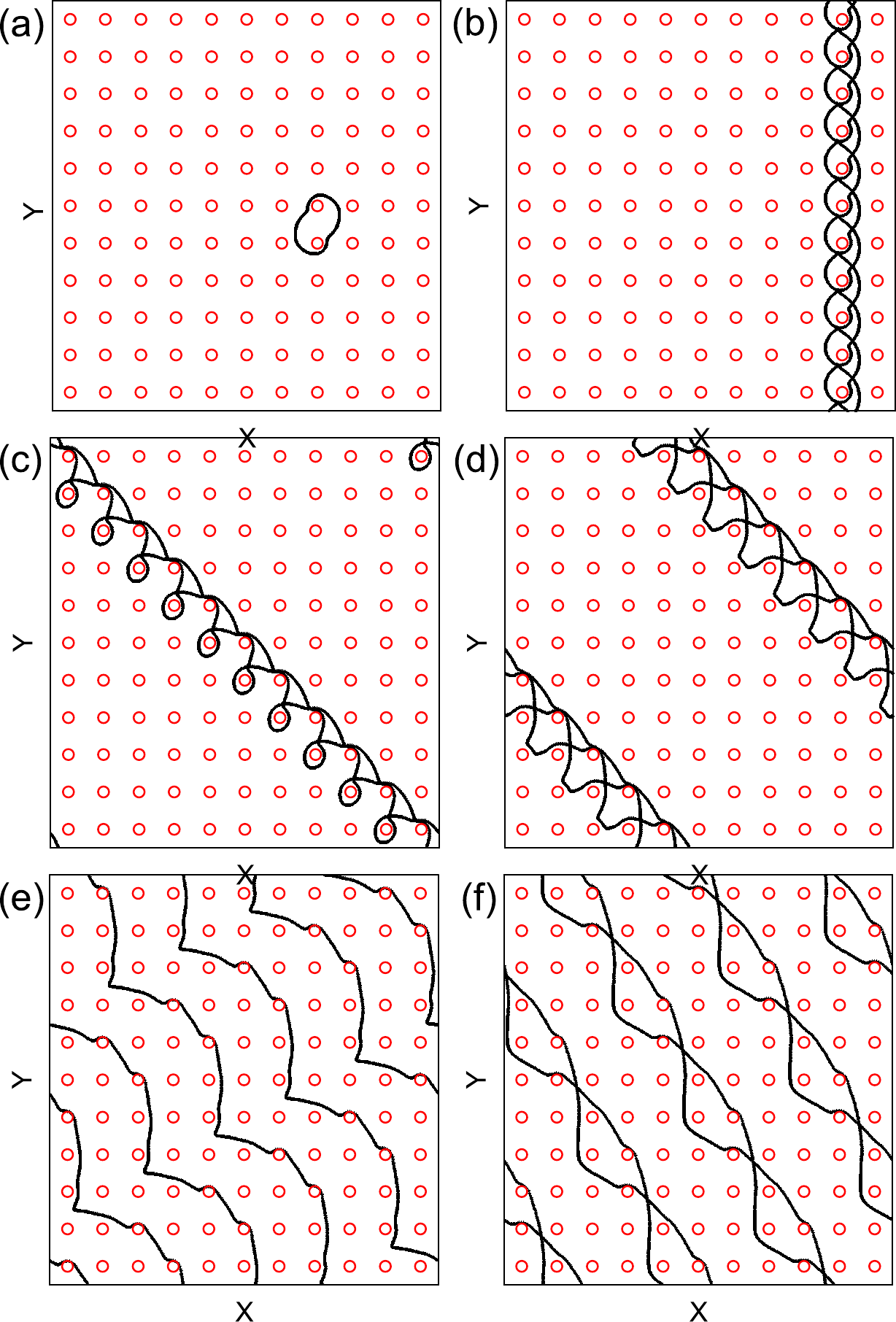}
\caption{ The obstacle locations (open circles) and skyrmion trajectory (line)
  for the system in Fig.~\ref{fig:5}
  with $\alpha_{m}/\alpha_{d} = 1.732$, $\omega_1=\omega_2$,
  and
  $A = B = 0.5$. 
  (a) $F^{D} = 0.011$ in the pinned phase.
  (b) Absolute transverse mobility at $F^{D} = 0.067$, where the skyrmion moves only
perpendicular to the dc driving direction.
(c) $F^{D} = 0.13$, where the motion is locked to $-45^\circ$. 
(d) $F^{D} = 0.47$. (e) $F^{D} =  0.62$. (f) $F^{D} = 0.95$.}  
\label{fig:6}
\end{figure}

In Fig.~\ref{fig:5}(a) we plot $\langle V_{||}\rangle$ and 
$\langle V_{\perp}\rangle$ versus $F^{D}$ for
a system with $\alpha_{m}/\alpha_{d} = 1.732$ and $A = B = 0.5$, while
in Fig.~\ref{fig:5}(b) we show the corresponding $\theta_{sk}$ versus $F^{D}$ curve.
Here, $\theta^{int}_{sk} = -60^\circ$.
The number of locking steps is higher
than for samples with
lower intrinsic skyrmion Hall angles, and
we observe a series of jumps
in $\theta_{sk}$.  The jumps primarily appear
for $F^{D} < 1.25$, while $\theta_{sk}$ has a more oscillating behavior for $F^{D} \geq 1.25$.
The inset of
Fig.~\ref{fig:5}(b) shows $\theta_{sk}$ over the range
$ 1.6 < F^{D} < 1.875$,
highlighting the large number of locking steps
that accompany a decrease in the magnitude of
$\theta_{sk}$.
As $F_{D}$ is increased further,
$\theta_{sk}$ gradually approaches the intrinsic value $\theta^{int}_{sk}$.
In Fig.~\ref{fig:6}(a) we plot the skyrmion trajectory for the system in
Fig.~\ref{fig:5} at $F^{D} = 0.011$, where a pinned orbit occurs in which
the skyrmion encircles two obstacles during each ac drive cycle.
In Fig.~\ref{fig:6}(b)
at $F^{D} = 0.067$,
$\langle V_{||}\rangle = 0.0$ and
$\langle V_{\perp}\rangle$ is finite,
giving $\theta_{sk} = -90^\circ$. This is an example of absolute transverse mobility 
in which the skyrmion translation is
strictly perpendicular to the applied dc drive.
The interval of $F^{D}$ over which transverse mobility occurs
is small, but it can be extended by varying other parameters as we demonstrate later.
At $F^D=0.13$ in
Fig.~\ref{fig:6}(c),
the motion is locked to $\theta_{sk} = -45^\circ$
and the skyrmion completes a loop
around  an obstacle during every ac cycle.
In Fig.~\ref{fig:6}(d)
at $F^{D} = 0.47$, the trajectory is still locked
to $\theta_{sk} = -45^\circ$ but the shape of the orbit has changed.
At $F^D=0.62$ in
Fig.~\ref{fig:6}(e),
$\theta_{sk} = -53.13^\circ$
and the skyrmion
translates a distance of $4a$ in the $y$ direction and $3a$ in the $x$
direction during each ac drive cycle.
In Fig.~\ref{fig:6}(f), the trajectories at $F^{D} = 0.95$ are one example
of the many possible phase locking orbits that the system exhibits.

In general, the locking steps
arise due to a combination of effects.
The first
is the directional locking associated with
the drive dependence
of the skyrmion Hall angle that occurs in the presence of pinning and in
the absence of ac driving
\cite{Reichhardt15a}.
Steps in $\theta_{sk}$ occur at prominent locking angles that match the symmetry
directions of the underlying obstacle lattice.
The second effect is the Shapiro or phase locking steps
that appear due to the
locking of the ac drive frequency with the
periodicity of the velocity component induced by the motion of the skyrmion
over the periodic substrate
\cite{Reichhardt15b}.
Since the skyrmion is moving in both the $x$ and $y$ directions,
there are two different velocity frequencies that can resonate with
the ac drive frequency in order to
create the Shapiro steps,
providing additional possible ways in which phase locking can occur.
The combined effect of the directional locking and the Shapiro steps
accounts for
the large number of phase locking steps that appear
in the velocity-force curves under combined dc and ac driving.
Many of these different phase locking effects compete with  one another,
producing frustration effects where, under certain driving conditions,
the skyrmion can lock to multiple phase locked orbits for the same $F^D$,
causing the skyrmion to
jump between the different orbits
and producing the chaotic regimes in the velocity-force curves.

\begin{figure}
\includegraphics[width=3.5in]{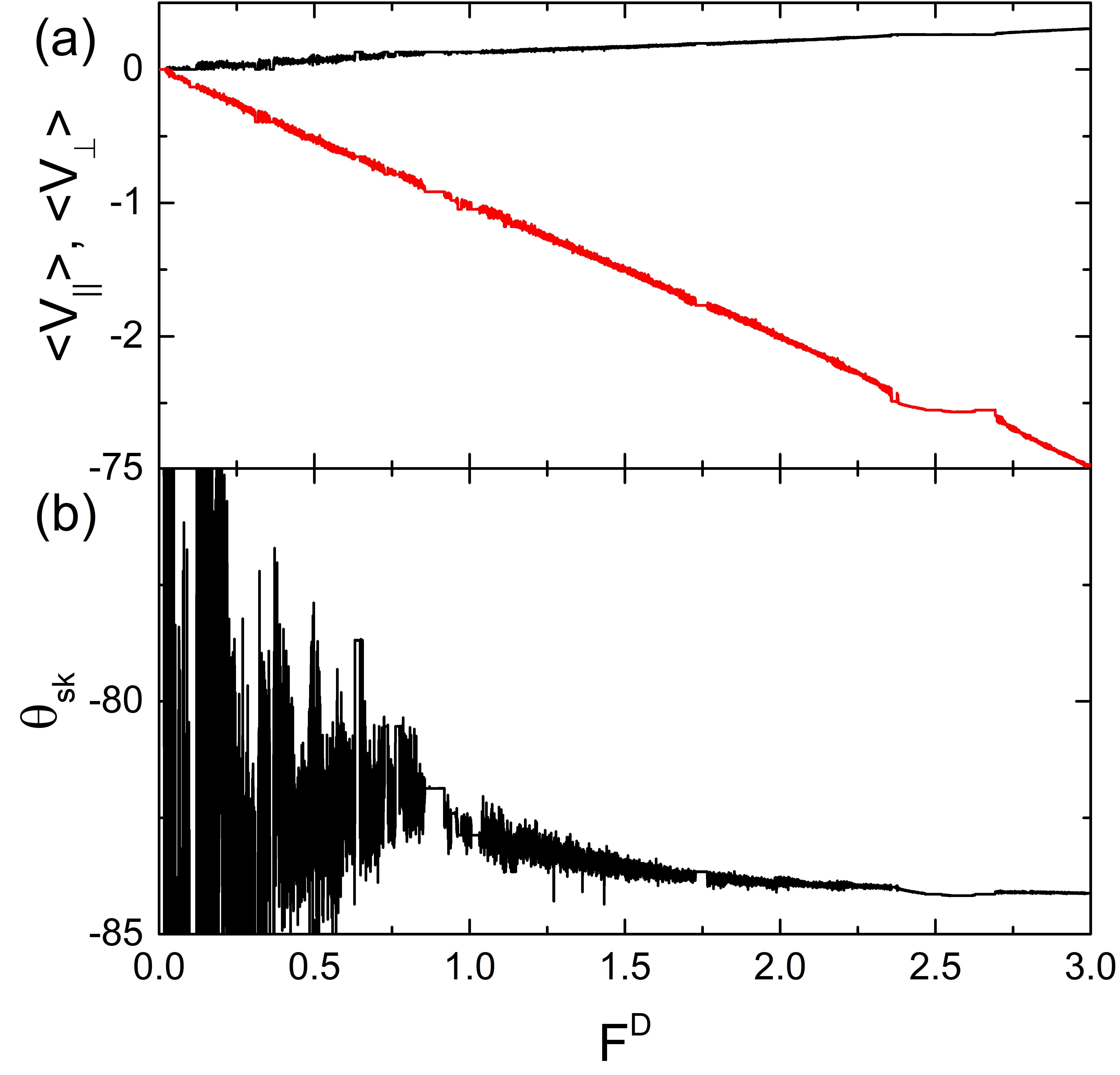}
\caption{
  (a) $\langle V_{\perp}\rangle$ (red) and $\langle V_{||}\rangle$ (black)
  vs $F^{D}$ for a system with $\alpha_{m}/\alpha_{d} = 9.962$,
  $\omega_1=\omega_2$,
  and
  $A = B = 0.5$.
  (b) The corresponding $\theta_{sk}$ vs $F^{D}$.} 
\label{fig:7}
\end{figure}

\begin{figure}
\includegraphics[width=3.5in]{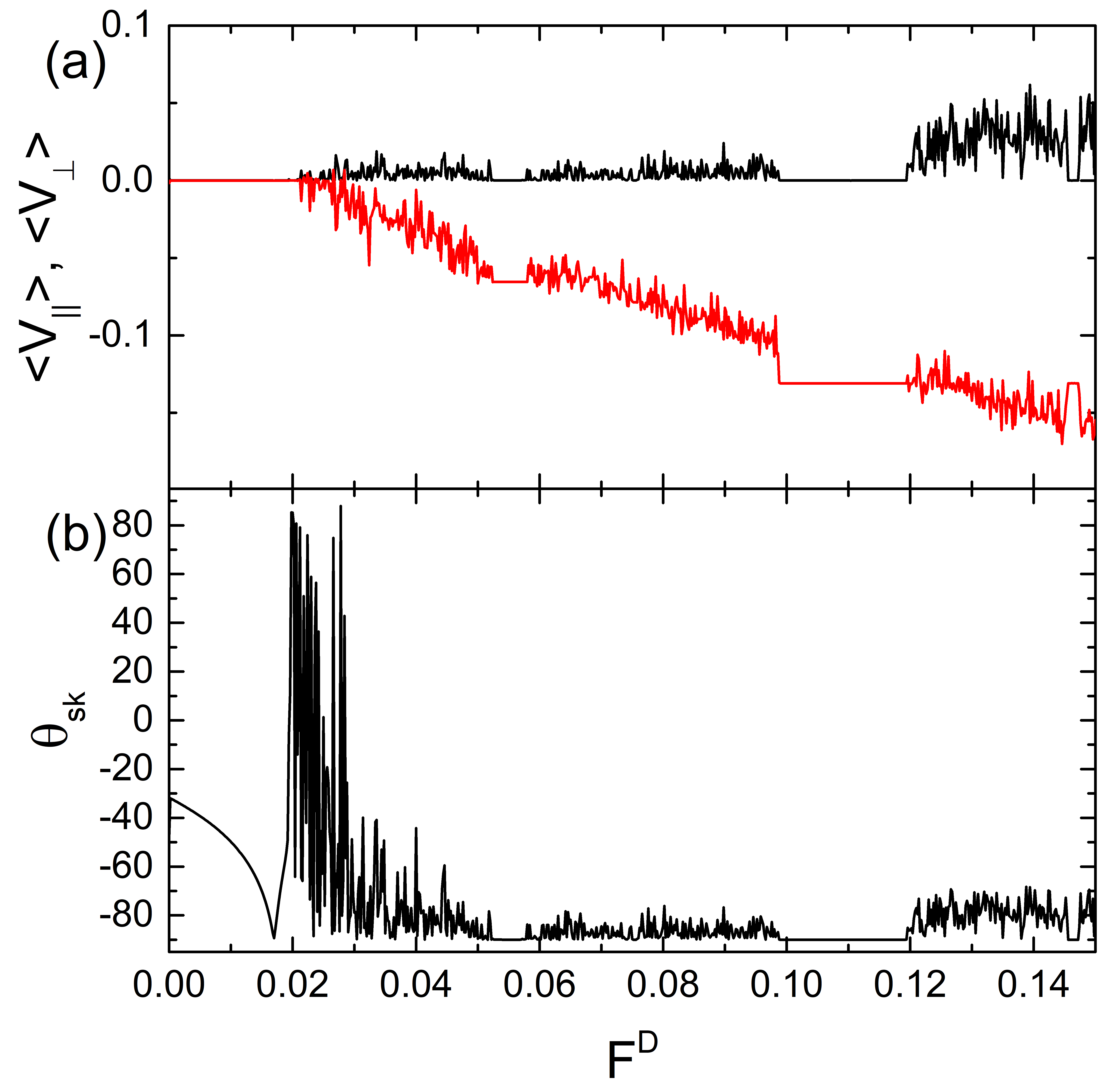}
\caption{A blowup of Fig.~\ref{fig:7} over the range
  $ 0 \leq F_{D} \leq 0.15$ for a system with
  $\alpha_{m}/\alpha_{d} = 9.962$,
  $\omega_1=\omega_2$,
  and
  $A = B = 0.5$.
  (a) $\langle V_{\perp}\rangle$ (red) and $\langle V_{||}\rangle$ (black)
  vs $F^{D}$.
  (b) $\theta_{sk}$ vs $F^{D}$.
  There is a pinned interval as well as an interval over which
  $\langle V_{||}\rangle = 0.0$ and $\langle V_{\perp}\rangle$ is finite,
  giving absolute transverse mobility.}
\label{fig:8}
\end{figure}

\begin{figure}
\includegraphics[width=3.5in]{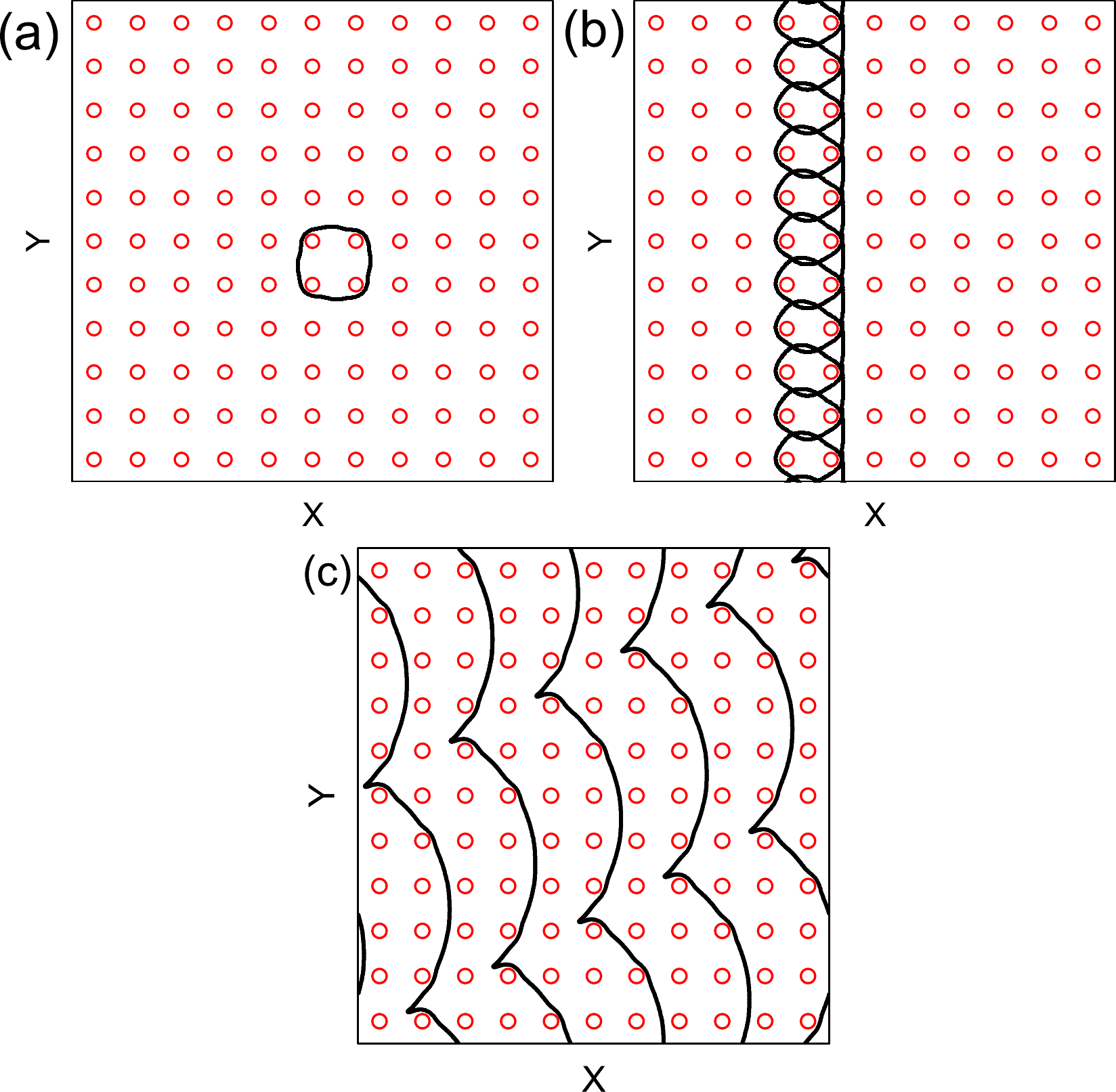}
\caption{The obstacle locations (open circles) and
  skyrmion trajectory (line) for the system 
in Figs.~\ref{fig:7} and \ref{fig:8} with 
  $\alpha_{m}/\alpha_{d} = 9.962$,
$\omega_1=\omega_2$,
and
  $A = B = 0.5$.
(a) $F^{D} = 0.01$, in the pinned phase. (b) $F^{D} = 0.11$,
where the skyrmion moves only perpendicular to the dc drive direction.
(c) $F^{D} = 0.63$, where there is motion in both the $x$ and $y$-directions.} 
\label{fig:9}
\end{figure}

In Fig.~\ref{fig:7}(a) we plot $\langle V_{\perp}\rangle$  and
$\langle V_{||}\rangle$ versus $F^D$ and in
Fig.~\ref{fig:7}(b) we show the corresponding $\theta_{sk}$
versus $F^D$ for a system with 
$\alpha_{m}/\alpha_{d} = 9.962$ and $A = B = 0.5$, where
$\theta^{int}_{sk} = -85.267^{\circ}$.
The windows of disordered motion are now larger,
but there are still some steps in the velocity-force curves
corresponding to different locking phases.
In Fig.~\ref{fig:8} we show a blow up of the 
velocity-force curves from Fig.~\ref{fig:7}
over the range $0 \leq F_{D} \leq 0.15$.
At low $F^D$, there is a pinned region with
$\langle V_{||}\rangle=\langle V_{\perp}\rangle=0$.
For higher $F^D$, we find a regime in which
$\langle V_{||}\rangle = 0$
while the magnitude of $\langle V_{\perp}\rangle$ is
increasing with increasing $F^D$, which is
an example of
transverse mobility.
In Fig.~\ref{fig:8}(b), the corresponding $\theta_{sk}$ versus $F^{D}$ 
shows that there is an interval over which the Hall angle is close to
$\theta_{sk}=-90^\circ$. 
In Fig.~\ref{fig:9}(a) we show the skyrmion trajectory
 for the system in Fig.~\ref{fig:8} at $F^{D} = 0.01$ in the pinned phase, 
 where the skyrmion  encircles four obstacles in a single ac drive cycle.
The radius of the pinned orbit increases with increasing Magnus force because
the Magnus term effectively magnifies the ac driving amplitude.
 At $F^D=0.11$ in Fig.~\ref{fig:9}(b),
 we illustrate
 the transverse mobility regime
 where the skyrmion moves in the negative $y$ direction
 and  encircles two obstacles during every ac drive cycle. 
In Fig.~\ref{fig:9}(c) at $F^{D} = 0.63$, the motion is locked to $\theta_{sk} = -78.60^\circ$ 
and the skyrmion translates a distance $a$ in the $x$ direction and
$5a$ in the $y$ direction during every ac drive cycle, giving $R = 5$.

\section{Varied ac Drive Amplitude}

\begin{figure}
\includegraphics[width=3.5in]{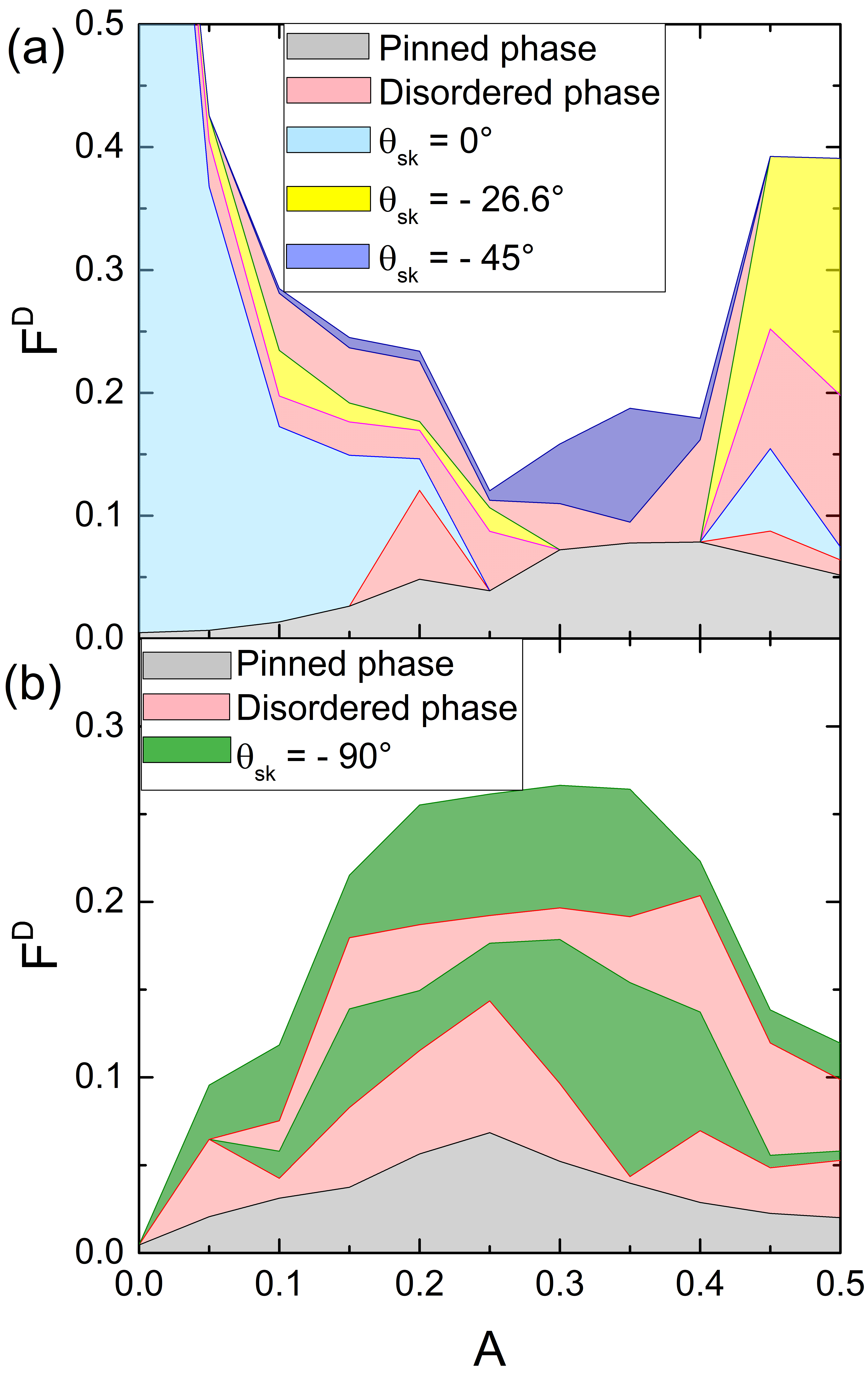}
\caption{
  Dynamic phase diagrams as a function of dc drive $F^D$ versus ac drive $A$ in systems
  with $\omega_1=\omega_2$
  and $A=B$.
  Pinned phase: gray; disordered phase: pink.
  (a) The system from Fig.~\ref{fig:2} with  $\alpha_m/\alpha_d=0.577$.
  Colors indicate the locking angles: $\theta_{sk}=0^\circ$ (blue),
  $-26.6^\circ$ (yellow), and $-45^\circ$ (purple).
  (b) The system from Figs.~\ref{fig:7} and \ref{fig:8} with
  $\alpha_m/\alpha_d=9.962$.
  Regions of locking to $\theta_{sk}=-90^\circ$ are colored in green.
}
\label{fig:10}
\end{figure}

We next consider the effect
on the velocity-force curves
of varying the ac drive amplitude
over the range $A=0.0$ to $A=0.5$ for a fixed driving frequency.
Due to the very large number of different locking effects that arise,
we summarize the results for
only a few selected locking phases and disordered phases.
In Fig.~\ref{fig:10}(a) we plot a dynamic phase diagram
as a function of dc drive $F^{D}$ versus ac drive amplitude $A$ 
for the system in Fig.~\ref{fig:2} with $\alpha_{m}/\alpha_{d} = 0.577$
and $A=B$.
We highlight only
the pinned phase,
disordered phase,
$\theta_{sk} = 0.0^{\circ}$ locking,
$\theta_{sk} = -26.6^\circ$ locking,
and $\theta_{sk} = -45^\circ$ locking.
When $A = 0$, the system is in the pinned phase
for $F^D<0.05$ and the
$\theta = 0.0^\circ$ phase for $0.05 \leq F^{D} < 0.5$.
As $F^D$ increases above $F^D=0.5$,
the system jumps to different locking phases (not shown).
When $A$ increases, the width of the pinned phase grows
until it reaches
a maximum near $A = 0.35$.
The $\theta = 0.0^\circ$ locking phase is absent
for $0.25 < A < 0.4$, which coincides with the window in which
the $\theta_{sk} = -45^\circ$ locking phase reaches its largest extent.
Disordered regions  appear
between the different locking phases.
For $A < 0.35$, the skyrmion  orbit is small enough that
the skyrmion is not able to encircle any obstacles,
while for $A > 0.35$, the orbit begins to encircle a single obstacle.
For $A > 0.5$ (not shown), we find a series of pinned phases
in which the skyrmion encircles
one, two, four, and then nine obstacles. 

In Fig.~\ref{fig:10}(b) we plot the dynamic phase diagram as a
function of $F^{D}$ versus $A$
for the system in Figs.~\ref{fig:7} and \ref{fig:8} with
$\alpha_{m}/\alpha_{d} = 9.962$.
We highlight only the pinned phase, the disordered phase, and the regime
of absolute transverse mobility with $\theta_{sk} = -90^\circ$.
The maximum extent of the pinned phase occurs for
$A = 0.25$, which also corresponds to the ac drive at which the
transverse mobility reaches its largest extent.
There are two distinct windows
of traverse mobility
that are associated with two different types of skyrmion orbits.
When $A<0.1$,
the skyrmion does not encircle
any obstacles,
while for larger $A$ it encircles four obstacles.
The reduction in the extent of the pinning and transverse mobility regions
for larger $A$
is the result of the larger orbit generated by the ac driving,
with the skyrmion jumping
to an orbit that encircles $9$ obstacles for $A > 0.5$.  

These results indicate that the transverse mobility is most prominent for
higher values of $\alpha_{m}/\alpha_{d}$ where the intrinsic Hall angle is the largest.

\begin{figure}
\includegraphics[width=3.5in]{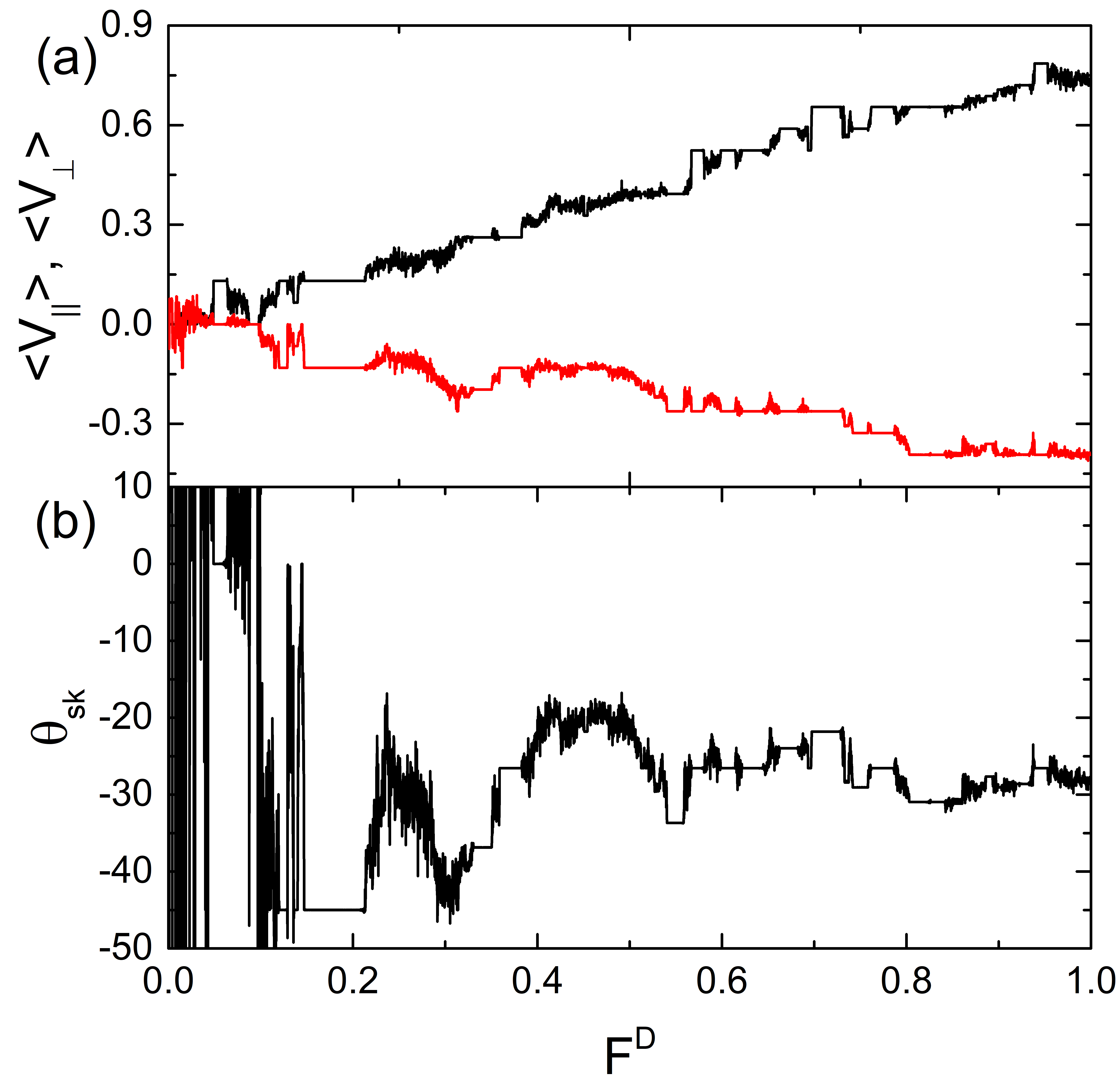}
\caption{ (a) $\langle V_{\perp}\rangle$ (red) and $\langle V_{||}\rangle$ (black)
  vs $F^{D}$ for a system with $\alpha_{m}/\alpha_{d} = 0.577$,
  $\omega_1=\omega_2$,
  $A = 0.5$, and $B = 1.0$.
  (b) The corresponding $\theta_{sk}$ vs $F^{D}$.} 
\label{fig:11}
\end{figure}

\begin{figure}
\includegraphics[width=3.5in]{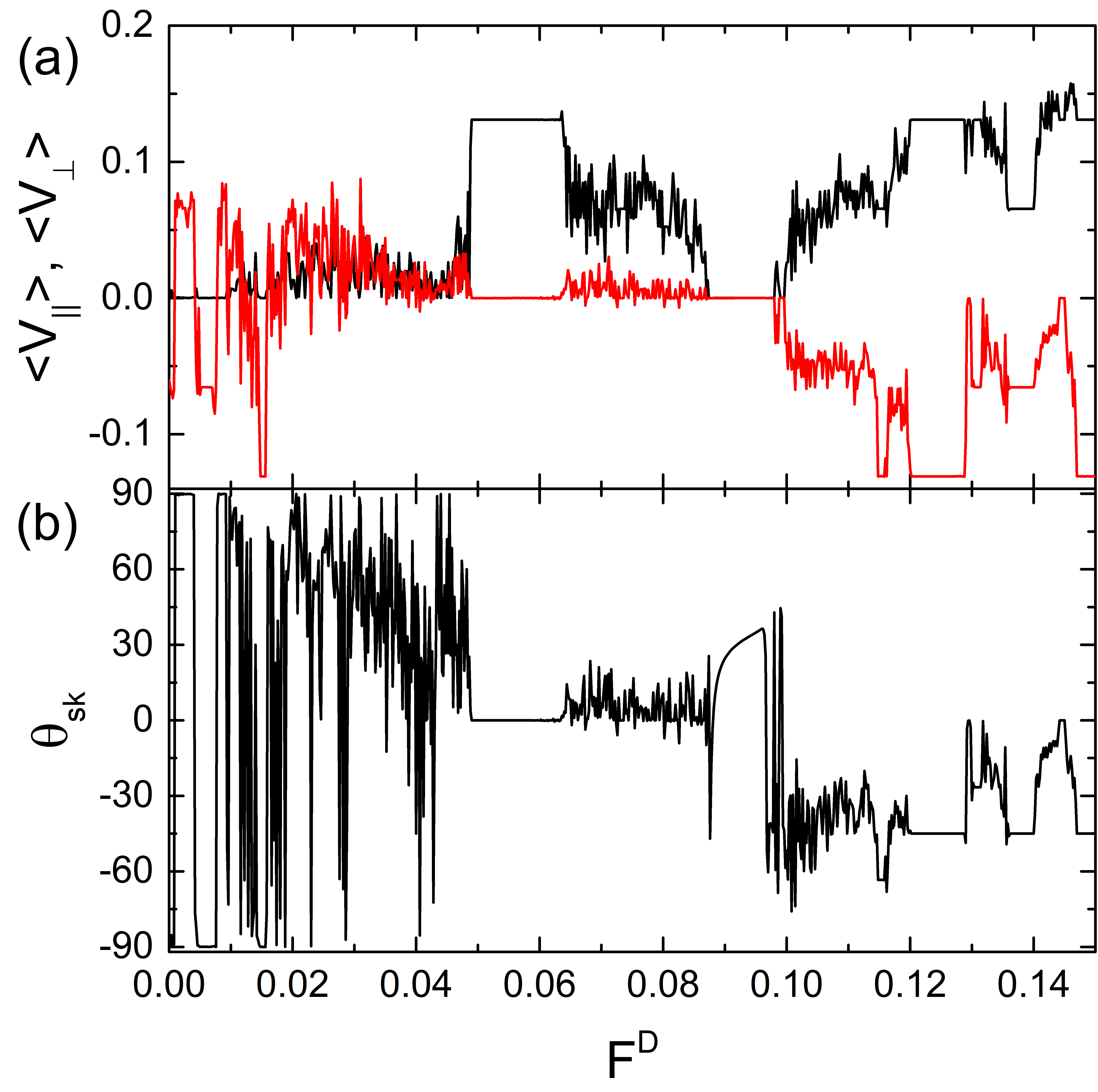}
\caption{ (a) $\langle V_{\perp}\rangle$ (red) and $\langle V_{||}\rangle$ (black)
  vs $F^{D}$ over the range $0 \leq F^D \leq 0.15$ for the system in Fig.~\ref{fig:11}
  with $\alpha_{m}/\alpha_{d} = 0.577$,
  $\omega_1=\omega_2$,
$A = 0.5$, and $B = 1.0$. (b) The corresponding $\theta_{sk}$ vs $F^{D}$.} 
\label{fig:12}
\end{figure}

\begin{figure}
  \begin{minipage}[c]{3.5in}
    \includegraphics[width=\textwidth]{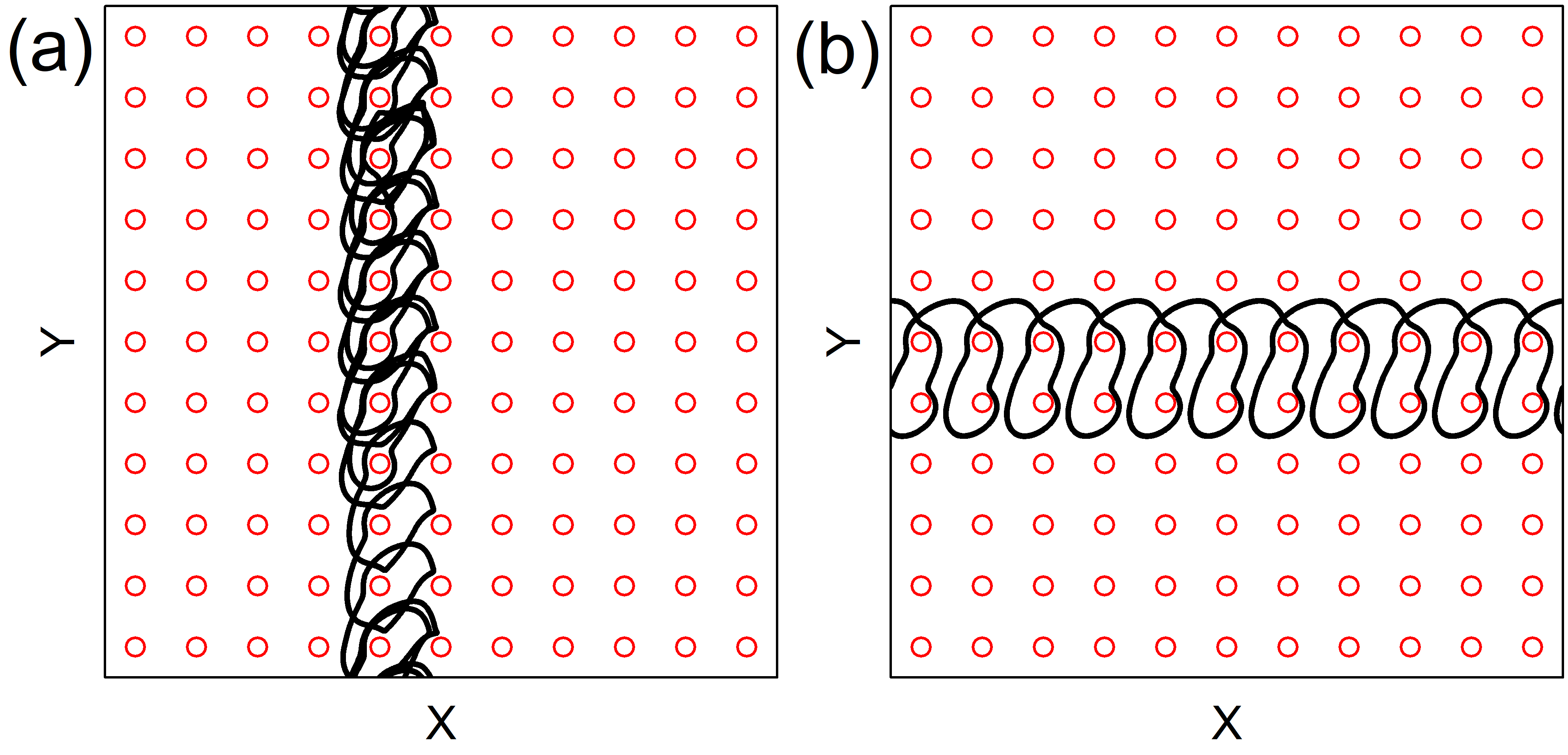}
  \end{minipage}
  \begin{minipage}[c]{3.5in}
    \includegraphics[width=\textwidth]{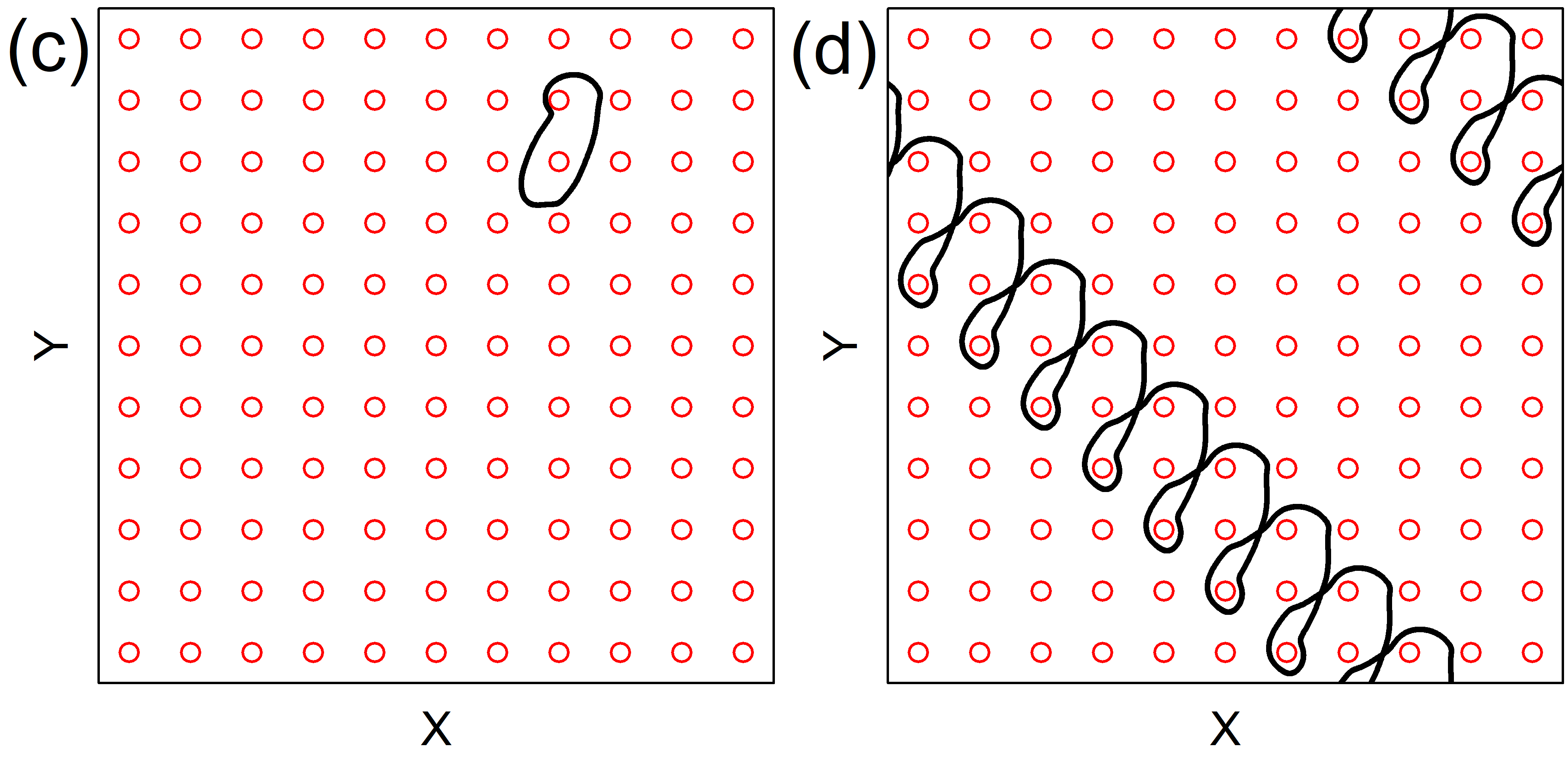}
    \end{minipage}
  \caption{The obstacle locations (open circles) and skyrmion trajectory (lines)
    for the system in Figs.~\ref{fig:11} and \ref{fig:12} with
  $\alpha_{m}/\alpha_{d} = 0.577$,
    $\omega_1=\omega_2$,
$A = 0.5$, and $B = 1.0$. 
(a) At $F^{D} = 0.008$, transverse mobility occurs where the skyrmion moves in the
    positive $y$ direction.
    (b) At $F^{D} = 0.055$, the motion is locked
in the $x$-direction. 
(c) $F^{D} = 0.093$ in the pinned phase.
(d) $F^{D}=0.2$, where the motion is locked to $\theta_{sk} = -45^\circ$.}  
\label{fig:13}
\end{figure}

\subsection{Two Different ac Amplitudes and Skyrmion Hall Angle Reversal}
We next consider the case where the ac drive amplitudes are different
in the two directions, $A \neq B$. 
In Fig.~\ref{fig:11} we plot $\langle V_{||}\rangle$,
$\langle V_{\perp}\rangle$, and $\theta_{sk}$ versus $F^{D}$
for a system with $\alpha_{m}/\alpha_{d} = 0.577$,
$A = 0.5$, and $B = 1.0$.
Here there is an extended region over which the system 
locks to $\theta_{sk}=-45^\circ$ followed by a gradual decline
to $\theta_{sk}=-30^\circ$ for higher drives
while a variety of locking steps and disordered regions appear.
In Fig.~\ref{fig:12}(a) we show
$\langle V_{||}\rangle$ and $\langle V_{\perp}\rangle$ versus $F^{D}$
for the same system as in Fig.~\ref{fig:11} but zoomed in over the range
$0.0 \leq F^{D} \leq 0.15$.
At low $F^{D}$, there is an extended region over which the motion 
is locked to the $x$-direction.
The system has a reentrant pinned region near $F^{D} = 0.9$ where
$\langle V_{\perp}\rangle=\langle V_{||}\rangle=0$.
Figure~\ref{fig:12}(b) 
shows the corresponding $\theta_{sk}$ versus $F^D$ curve.
For $F^{D} < 0.045$, the skyrmion Hall angle is oscillatory and
undergoes repeated reversals from positive to negative values.
The finite value of $\theta_{sk}$ in the pinned region near $F^D=0.9$ results
from the undefined $\theta_{sk}$ calculation that occurs when
both the parallel and perpendicular
velocities are zero.
In Fig.~\ref{fig:13}(a) we illustrate the skyrmion trajectories for the system in
Fig.~\ref{fig:12} at $F^{D} = 0.008$ where
transverse mobility occurs.
The skyrmion is moving in the positive $y$ direction, giving a positive
skyrmion Hall angle.
In Fig.~\ref{fig:13}(b) at $F^{D} = 0.055$, the motion is locked in the $x$-direction
and the skyrmion encircles two obstacles during every ac drive cycle.
Figure~\ref{fig:13}(c) shows the pinned orbit at $F^{D} = 0.093$, 
where the skyrmion encircles two obstacles but does not translate.
In Fig.~\ref{fig:13}(d) at $F^{D} = 0.2$, the
motion is locked to $\theta_{sk} = -45^\circ$.

\begin{figure}
\includegraphics[width=3.5in]{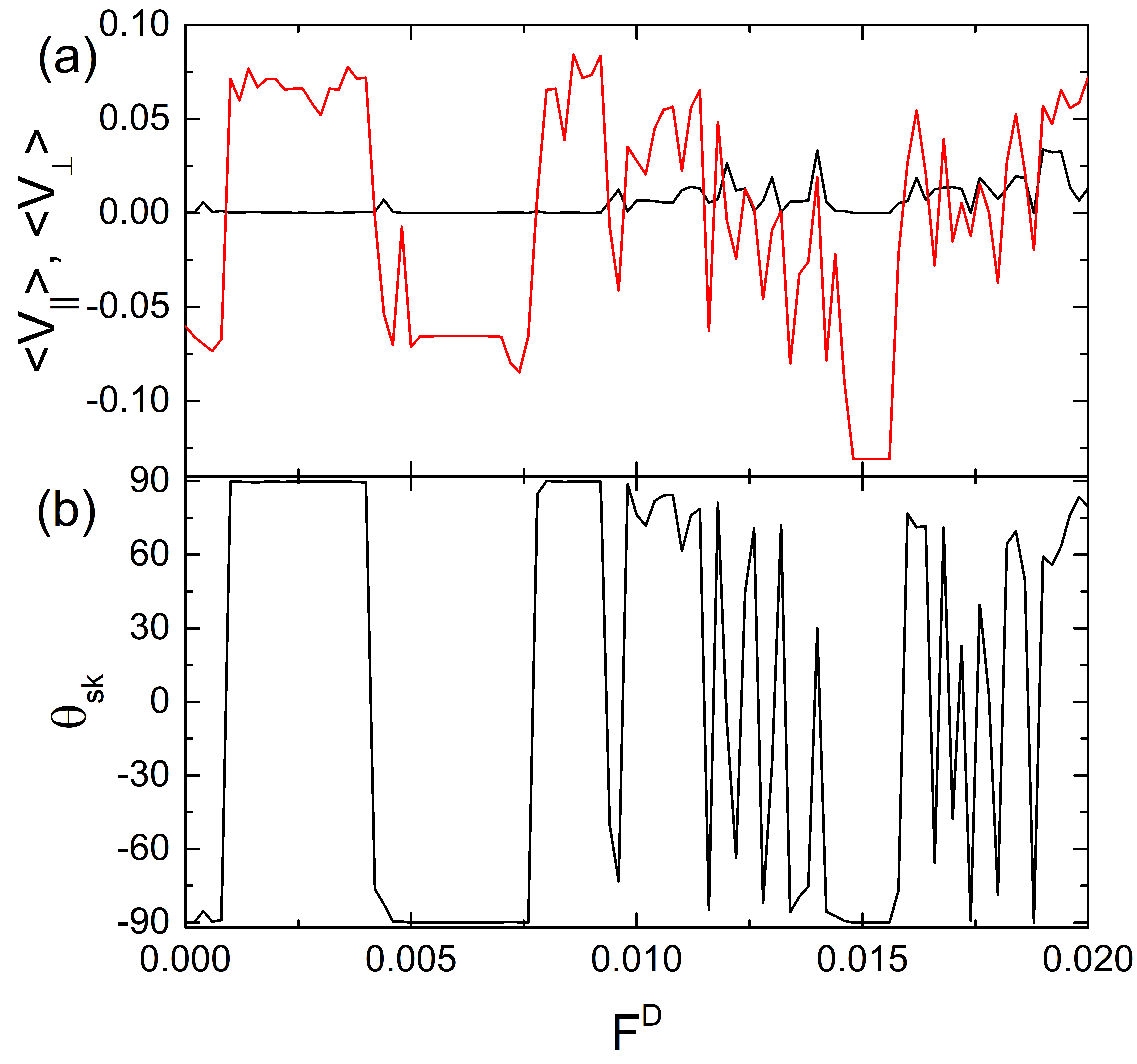}
\caption{
  (a) $\langle V_{||}\rangle$ (black) and $\langle V_{\perp}\rangle$ vs
  $F^{D}$ for the system in Figs.~\ref{fig:11} and \ref{fig:12}
  with $\alpha_{m}/\alpha_{d} = 0.577$,
  $\omega_1=\omega_2$,
$A = 0.5$, and $B = 1.0$ 
  for $F^{D} \leq 0.02$, 
  showing the reversal in the transverse mobility.
  (b) The corresponding $\theta_{sk}$ vs $F^{D}$.}  
\label{fig:14}
\end{figure}

In Fig.~\ref{fig:14}(a) we show a zoom of $\langle V_{||}\rangle$
and $\langle V_{\perp}\rangle$ versus $F^{D}$ for the system in
Fig.~\ref{fig:12}
over the range $0 \leq F^{D} \leq 0.02$,
and in Fig.~\ref{fig:14}(b) we show the corresponding $\theta_{sk}$.
The system passes though a series of locked phases
that are associated with transverse mobility, 
but there are also
repeated reversals of the Hall angle with
increasing $F^{D}$.
After each reversal, the system locks to a different orbit.
Another interesting feature is that at 
$F^{D} = 0.0$, the skyrmion has a finite velocity
in the negative $y$-direction. 
This motion, which occurs under only the ac drive without a dc drive,
represents a type of ratchet effect.
In overdamped systems, similar ratchet effects can occur
for a particle on a periodic substrate 
that is subjected to biharmonic ac drives \cite{Reichhardt03,Speer09}.
The ratchet effect occurs when enough symmetries are broken in a nonequilibrium
system.
The circular ac drive breaks a chiral symmetry,
but in the absence of a substrate asymmetry,
the ac orbit must itself be spatially
asymmetric in order to produce the ratchet effect.
A simple circular ac drive with $A=B$ does not give rise to a ratchet effect.
In the skyrmion system,
the Magnus force combined with the dc driving can produce asymmetric orbits,
as illustrated for the pinned phase in Fig.~\ref{fig:13}(c).
The ratchet effects that occur at $F^{D} = 0.0$ will be explored
more fully in a future work.
In general, 
the ratchet effects are more relevant at lower $F^{D}$,
whereas for high dc drives the Shapiro step and directional locking effects
dominate the behavior.

\begin{figure}
\includegraphics[width=3.5in]{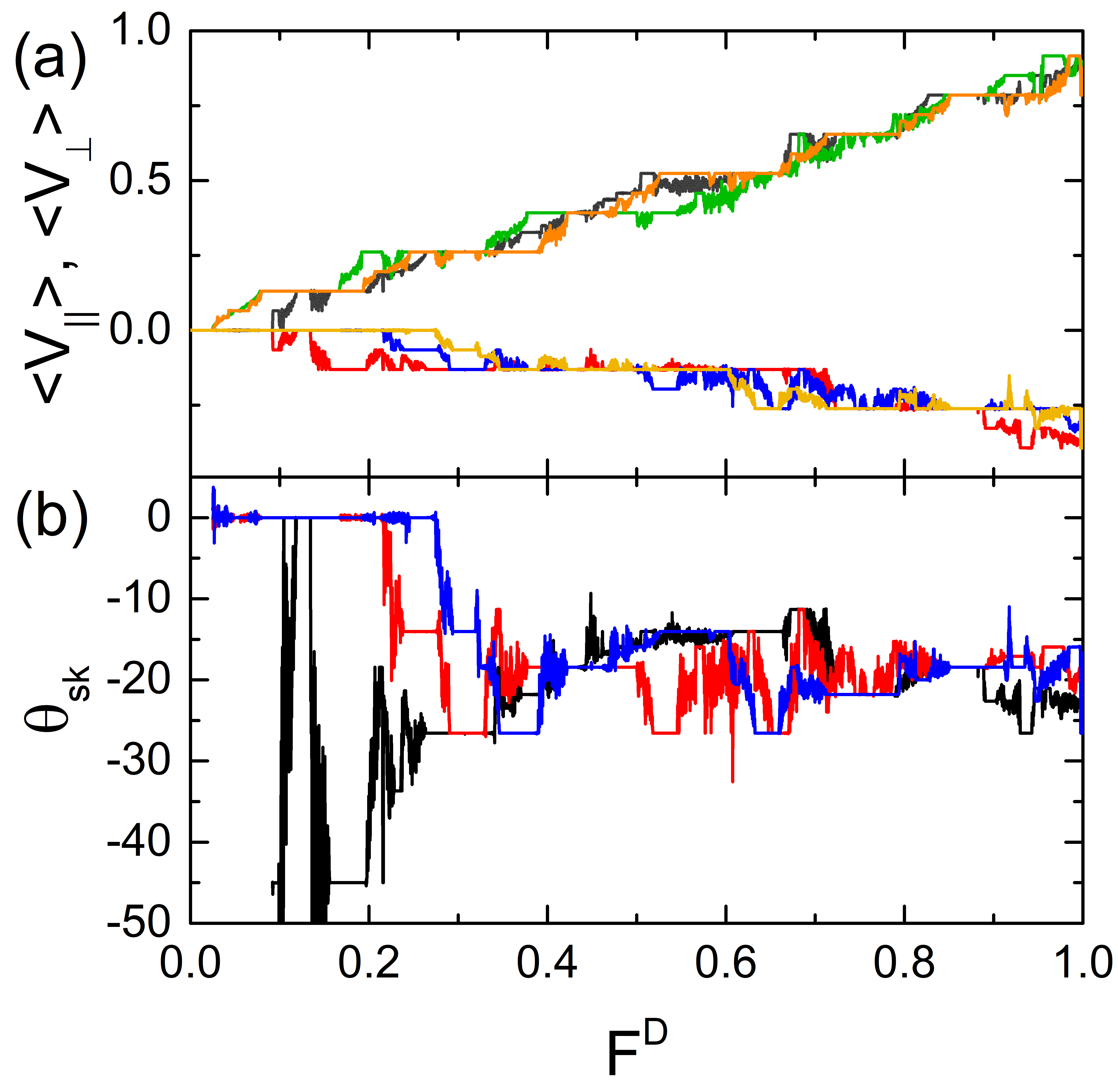}
\caption{
  (a) $\langle V_{||}\rangle$ and $\langle V_{\perp}\rangle$ vs $F^{D}$
  for a system with $A = B = 0.5$, $\omega_{1} = 2\times 10^{-4}$,
and $\alpha_{m}/\alpha_{d} = 0.45$. 
$\omega_{2} = 2\omega_{1}$ (black, red). 
$\omega_{2} = 3\omega_{3}$ (green, blue). 
$\omega_{2} = 4\omega_{3}$ (orange, yellow). 
(b) The corresponding $\theta_{sk}$ vs $F^D$.
$\omega_{2} = 2\omega_{1}$ (black).
$\omega_{2} = 3\omega_{3}$ (blue).  
$\omega_{2} = 4\omega_{3}$ (red).}  
\label{fig:15}
\end{figure}

\begin{figure}
\includegraphics[width=3.5in]{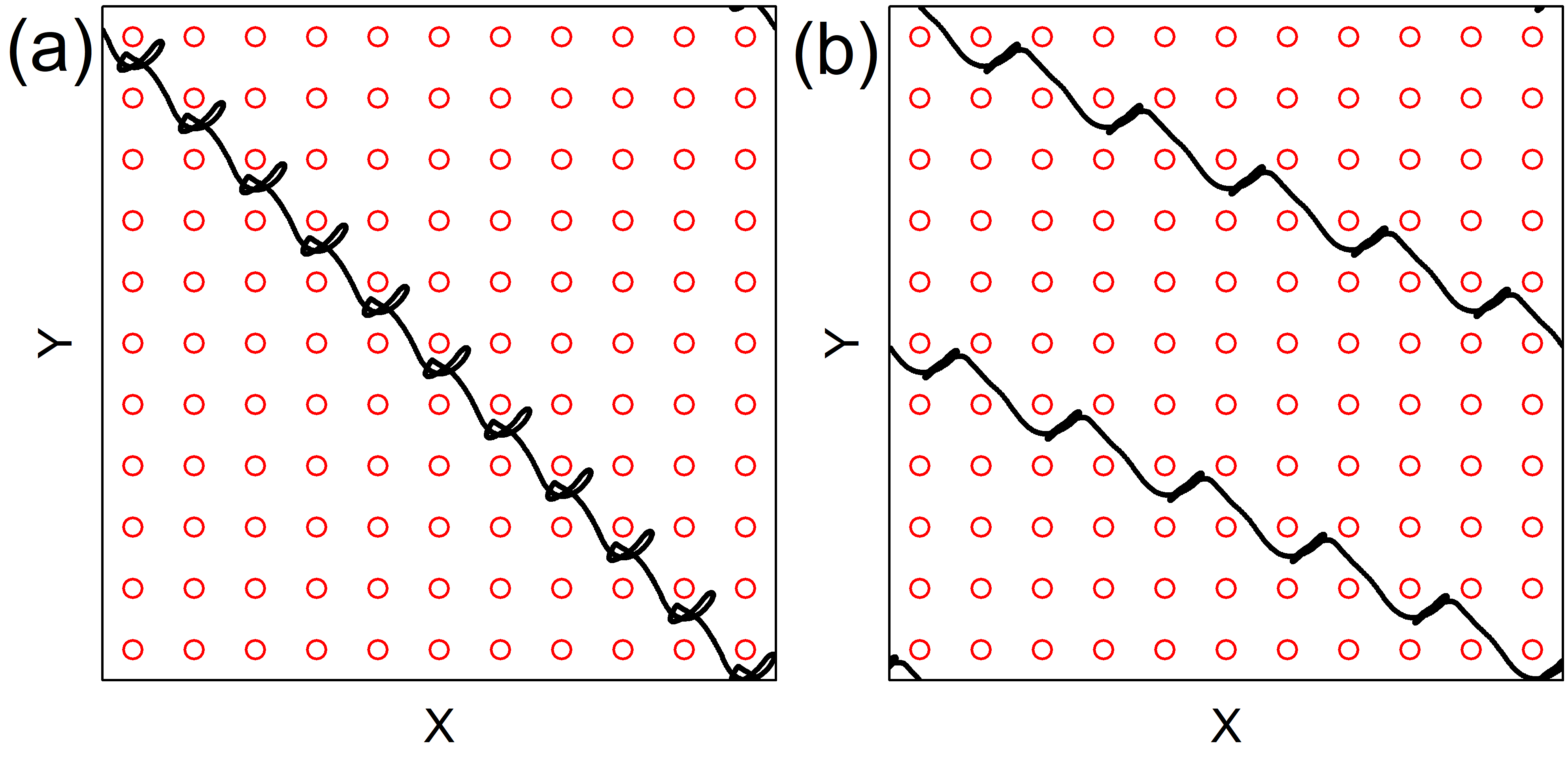}
\caption{
  The obstacle locations (open circles) and skyrmion trajectory (lines)
  for the $\omega_{2} = 2\omega_{1}$ sample from Fig.~\ref{fig:15},
  where
  $A = B = 0.5$, $\omega_{1} = 2\times 10^{-4}$,
  and $\alpha_{m}/\alpha_{d} = 0.45$.
  (a) At $F^{D} = 0.177$, the skyrmion is locked to
  $\theta_{sk}=-45^\circ$ and 
  performs a double loop during each ac drive cycle.
  (b) At $F^{D} = 0.3$, the skyrmion
is moving at $\theta_{sk} = -26.56^\circ$.} 
\label{fig:16}
\end{figure}

We next consider the effect of holding the ac drive amplitudes fixed at $A=B$
but varying the ac drive frequencies so that $\omega_1 \neq \omega_2$.
In Fig.~\ref{fig:15}(a), we plot $\langle V_{||}\rangle$ and
$\langle V_{\perp}\rangle$ for systems with
$\alpha_m/\alpha_d=0.45$, $A=B=0.5$ and $\omega_1=2\times 10^{-4}$ at
$\omega_{2} = 2\omega_{1}$,
$3\omega_{1}$,
and
$4\omega_{1}$,
while in Fig.~\ref{fig:15}(b) we show the corresponding $\theta_{sk}$ versus $F^D$
curves.
We observe several trends.
Certain locking phases occur for all three values of $\omega_2$;
however, the width of the locked phases varies as $\omega_2$ varies.
When $\omega_{2} = 2\omega_{1}$,
the system locks to $-45^{\circ}$ at lower drives and then gradually approaches the 
intrinsic Hall angle.
In Fig.~\ref{fig:16}(a) we illustrate the skyrmion trajectory for the
$\omega_{2} = 2\omega_{1}$ sample at
$F^{D} = 0.177$, where the skyrmion is locked to
$\theta_{sk}=-45^\circ$
and performs a double loop
during each ac drive cycle.
Figure~\ref{fig:16}(b) shows the trajectory at $F^{D} = 0.3$, where the skyrmion 
is moving at $\theta_{sk} = -26.56^\circ$.

\begin{figure}
\includegraphics[width=3.5in]{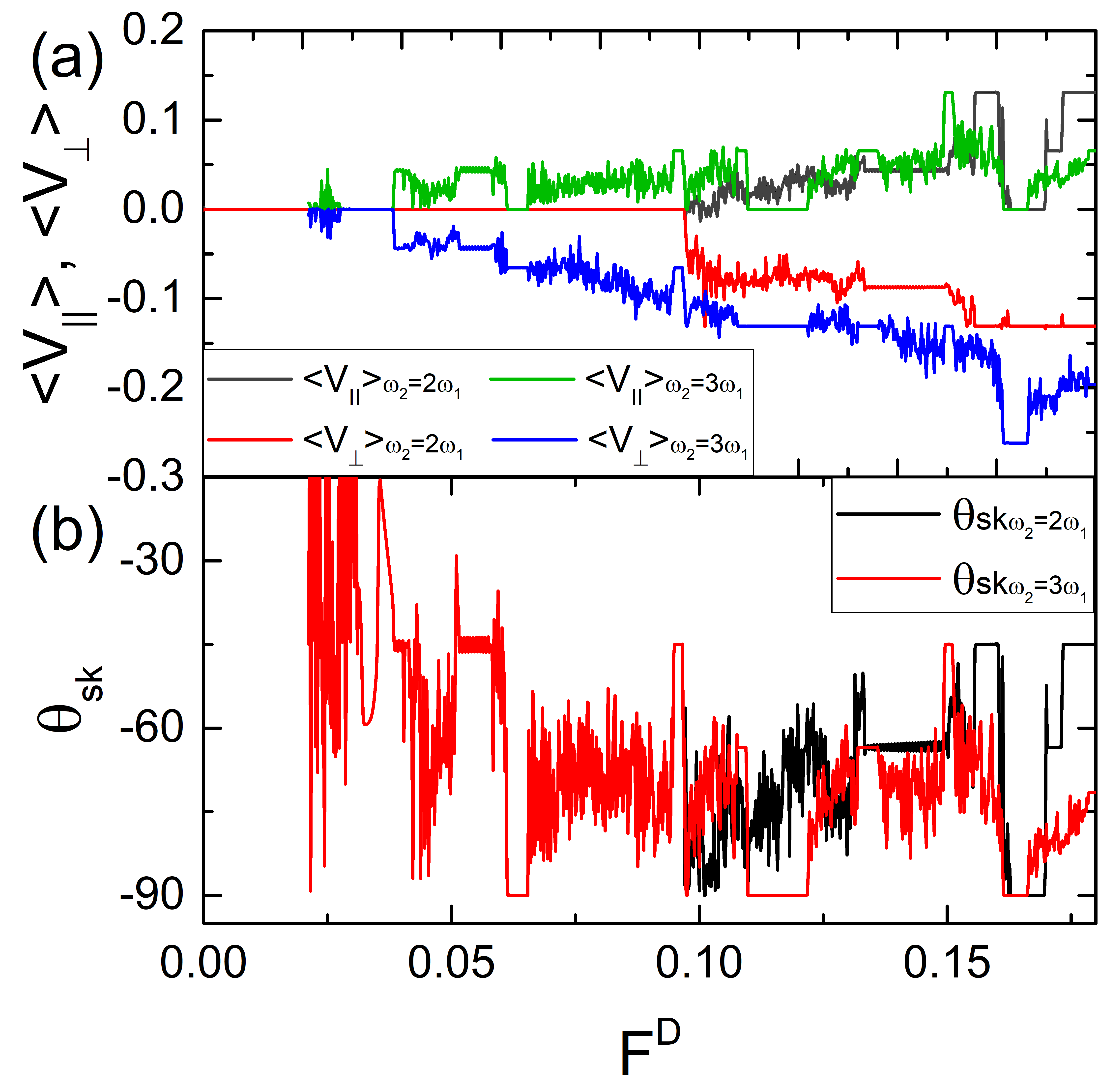}
\caption{
  (a) $\langle V_{||}\rangle$ and $\langle V_{\perp}\rangle$ vs $F^{D}$
  for a system with $A = B = 0.5$, $\omega_{1} = 2\times 10^{-4}$,
  and $\alpha_{m}/\alpha_{d} = 1.732$.
  $\omega_{2} = 2\omega_{1}$ (black, red). 
  $\omega_{2} = 3\omega_{3}$. (green, blue).
  (b) The corresponding $\theta_{sk}$ vs $F^{D}$. 
  $\omega_{2} = 2\omega_{1}$ (black).
  $\omega_{2} = 3\omega_{1}$ (red).}  
\label{fig:17}
\end{figure}

In Fig.~\ref{fig:17}(a) we plot $\langle V_{||}\rangle$ and
$\langle V_{\perp}\rangle$ versus $F^{D}$ for
a system with $A = B = 0.5$ and $\alpha_{m}/\alpha_{d} = 1.732$. 
When $\omega_{2} = 3\omega_{1}$,
there is an initial pinned phase at low $F^D$.
The system locks to $-45^\circ$ over several drive intervals,
and there are also several regions in which
absolute transverse mobility
occurs with $\theta_{sk} = -90^\circ$ as shown in the plot of
$\theta_{sk}$ versus $F^{D}$ in Fig.~\ref{fig:17}(b). 
When $\omega_{2} = 2\omega_{1}$,
the pinned region extends out to larger drives,
and there are also several drive intervals
at which $\theta_{sk} = -90^\circ$.
These results show that the transverse mobility can be
enhanced by varying the ac drive frequencies.  

\section{Discussion}
In this work, we neglected temperature;
however, thermal effects can be important
in certain skyrmion systems.
Thermal fluctuations can wash out directional phase locking,
but in some cases they can also induce
other types of phase locking effects \cite{Mali20}. 
Experiments that could be performed in this system include
direct imaging of skyrmions and
measurements of changes in the topological Hall effect.
Another route for further exploration
would be to examine noise fluctuations \cite{Diaz17}
in order to 
observe the emergence of
narrow band signals associated with phase locking.
It has already been shown experimentally
that such measurements are possible in skyrmion systems \cite{Sato19}.
In the locked phases, the skyrmion motion should
be periodic and produce a large narrow band noise signal,
while in the disordered regions this signal will be reduced or lost.
It would also be interesting to explore the effect of
the internal modes of the skyrmions \cite{Chen19,Chen20},
which could induce additional oscillating signals
that might produce different types of phase locking.
We have focused only on the case of a single skyrmion;
however, if lattices of skyrmions
interact with 2D periodic arrays,
we expect that additional collective effects
would occur that would depend on the filling factor
or the number of skyrmions relative to the number of pinning sites.
Slightly away from commensurate fillings, at which the number of skyrmions is an
integer multiple of the number of pinning sites,
soliton like states can appear which could themselves
exhibit Shapiro steps and other phase locking phenomena,
similar to what has been observed in colloidal and
superconducting vortex systems with periodic substrate arrays.
The additional Magnus force that is present in the skyrmion
system could induce new types of dynamics that do not occur in
overdamped systems.   

\section{Summary}
We examined a skyrmion interacting with a 2D periodic array of obstacles
under an applied dc drive and biharmonic ac drive,
and find a rich variety of nonlinear dynamical effects
due to the presence of the Magnus force and the
velocity dependence of the skyrmion Hall angle.
A biharmonic ac drive alone creates a circular skyrmion orbit in the absence of
obstacles or a dc drive.
Under only dc driving and in the presence of the periodic obstacles, the
skyrmion passes through a series of directional locking phases
due to the velocity dependence of the skyrmion Hall
effect.
When a finite biharmonic ac drive is included, we find that
the velocity-force curves show a series of jumps and locking intervals
in which
the skyrmion motion locks to specific symmetry directions of the substrate.
Within these locking phases,
the skyrmion can encircle multiple obstacles during each ac drive cycle.
We also observe regimes in which the skyrmion motion is disordered
and the motion is not locked to a fixed direction.  
We find that the locking phases
can be associated with both increases and
decreases in the skyrmion Hall angle.
Many of the locking
phases are reentrant and recur repeatedly
for increasing dc drive.
In general, as the Magnus force increases,
the skyrmion encircles a larger number of obstacles
during each ac drive cycle,
and for large Magnus forces, we observe
a series of absolute transverse mobility phases
in which the skyrmion moves at exactly $90^\circ$ with respect to the
dc driving direction.
We find reentrance in both the transverse mobility and the pinning phase.
We show that it is possible to have oscillations
in the transverse mobility where the skyrmion Hall angle
switches between $\theta_{sk}=-90^\circ$ and $\theta_{sk}=90^\circ$.
We also find that the
transverse mobility can be enhanced when the
two ac drives have different amplitudes or different frequencies
so that the driving is no longer circular.

%
%

This work was supported by the US Department of Energy through
the Los Alamos National Laboratory.  Los Alamos National Laboratory is
operated by Triad National Security, LLC, for the National Nuclear Security
Administration of the U. S. Department of Energy (Contract No. 892333218NCA000001).
N.P.V. acknowledges
funding from
Funda\c{c}\~{a}o de Amparo \`{a} Pesquisa do Estado de S\~{a}o Paulo - FAPESP (Grant 2018/13198-7).

\section{Authors contributions}
All the authors were involved in the preparation of the manuscript.
All the authors have read and approved the final manuscript.

\bibliographystyle{epj}
\bibliography{mybib}

\begin{thebibliography}{92}

\bibitem{Harada96}
K.~Harada, O.~Kamimura, H.~Kasai, T.~Matsuda, A.~Tonomura, V.V. Moshchalkov,
  Science \textbf{274}, 1167 (1996)

\bibitem{Reichhardt97}
C.~Reichhardt, C.J. Olson, F.~Nori, Phys. Rev. Lett. \textbf{78}, 2648 (1997)

\bibitem{Martin99}
J.I. Mart\'{\i}n, M.~V\'elez, A.~Hoffmann, I.K. Schuller, J.L. Vicent, Phys.
  Rev. Lett. \textbf{83}, 1022 (1999)

\bibitem{Reichhardt08a}
C.~Reichhardt, C.J.O. Reichhardt, Phys. Rev. B \textbf{78}, 224511 (2008)

\bibitem{Gutierrez09}
J.~Gutierrez, A.V. Silhanek, J.~Van~de Vondel, W.~Gillijns, V.V. Moshchalkov,
  Phys. Rev. B \textbf{80}, 140514 (2009)

\bibitem{Sadovskyy17}
I.A. Sadovskyy, Y.L. Wang, Z.L. Xiao, W.K. Kwok, A.~Glatz, Phys. Rev. B
  \textbf{95}, 075303 (2017)

\bibitem{Ge18}
J.Y. Ge, V.N. Gladilin, J.~Tempere, J.T. Devreese, V.V. Moshchalkov, Nature
  Commun. \textbf{9}, 2576 (2018)

\bibitem{Korda02}
P.T. Korda, M.B. Taylor, D.G. Grier, Phys. Rev. Lett. \textbf{89}, 128301
  (2002)

\bibitem{MacDonald03}
M.P. MacDonald, G.C. Spalding, K.~Dholakia, Nature (London) \textbf{426}, 421
  (2003)

\bibitem{Bohlein12}
T.~Bohlein, J.~Mikhael, C.~Bechinger, Nature Mater. \textbf{11}, 126 (2012)

\bibitem{Vanossi12}
A.~Vanossi, N.~Manini, E.~Tosatti, Proc. Natl. Acad. Sci. (USA) \textbf{109},
  16429 (2012)

\bibitem{Hasnain13}
J.~Hasnain, S.~Jungblut, C.~Dellago, Soft Matter \textbf{9}, 5867 (2013)

\bibitem{McDermott13a}
D.~McDermott, J.~Amelang, C.J.O. Reichhardt, C.~Reichhardt, Phys. Rev. E
  \textbf{88}, 062301 (2013)

\bibitem{Tierno09}
P.~Tierno, F.~Sagues, T.H. Johansen, T.M. Fischer, Phys. Chem. Chem. Phys.
  \textbf{11}, 9615 (2009)

\bibitem{Loehr18}
J.~Loehr, D.~de~las Heras, A.~Jarosz, M.~Urbaniak, F.~Stobiecki, A.~Tomita,
  R.~Huhnstock, I.~Koch, A.~Ehresmann, D.~Holzinger et~al., Commun. Phys.
  \textbf{1}, 4 (2018)

\bibitem{Cao19}
X.~Cao, E.~Panizon, A.~Vanossi, N.~Manini, C.~Bechinger, Nature Phys.
  \textbf{15}, 776 (2019)

\bibitem{Stoop20}
R.L. Stoop, A.V. Straube, T.H. Johansen, P.~Tierno, Phys. Rev. Lett.
  \textbf{124}, 058002 (2020)

\bibitem{Tekic05}
J.~Teki\ifmmode~\acute{c}\else \'{c}\fi{}, O.M. Braun, B.~Hu, Phys. Rev. E
  \textbf{71}, 026104 (2005)

\bibitem{Vanossi13}
A.~Vanossi, N.~Manini, M.~Urbakh, S.~Zapperi, E.~Tosatti, Rev. Mod. Phys.
  \textbf{85}, 529 (2013)

\bibitem{Shapiro63}
S.~Shapiro, Phys. Rev. Lett. \textbf{11}, 80 (1963)

\bibitem{Barone82}
A.~Barone, G.~Paterno, \emph{Physics and Applications of the {Josephson}
  effect} (Wiley, New York, 1982)

\bibitem{Martinoli75}
P.~Martinoli, O.~Daldini, C.~Leemann, E.~Stocker, Sol. St. Commun. \textbf{17},
  205 (1975)

\bibitem{vanLook99}
L.~Van~Look, E.~Rosseel, M.J. Van~Bael, K.~Temst, V.V. Moshchalkov,
  Y.~Bruynseraede, Phys. Rev. B \textbf{60}, R6998 (1999)

\bibitem{Reichhardt00b}
C.~Reichhardt, R.T. Scalettar, G.T. Zim\'anyi, N.~Gr\o{}nbech-Jensen, Phys.
  Rev. B \textbf{61}, R11914 (2000)

\bibitem{Dobrovolskiy15}
O.V. Dobrovolskiy, J. Supercond. Novel Mag. \textbf{28}, 469 (2015)

\bibitem{Juniper15}
M.P.N. Juniper, A.V. Straube, R.~Besseling, D.G.A.L. Aarts, R.P.A. Dullens,
  Nature Commun. \textbf{6}, 7187 (2015)

\bibitem{Brazda17}
T.~Brazda, C.~July, C.~Bechinger, Soft Matter \textbf{13}, 4024 (2017)

\bibitem{Reichhardt01}
C.~Reichhardt, A.B. Kolton, D.~Dom\'{\i}nguez, N.~Gr\o{}nbech-Jensen, Phys.
  Rev. B \textbf{64}, 134508 (2001)

\bibitem{Marconi03}
V.I. Marconi, A.B. Kolton, D.~Dom\'{\i}nguez, N.~Gr\o{}nbech-Jensen, Phys. Rev.
  B \textbf{68}, 104521 (2003)

\bibitem{Reichhardt02b}
C.~Reichhardt, C.J. Olson, Phys. Rev. B \textbf{65}, 100501 (2002)

\bibitem{Guantes03}
R.~Guantes, S.~Miret-Art\'es, Phys. Rev. E \textbf{67}, 046212 (2003)

\bibitem{Reichhardt03}
C.~Reichhardt, C.J. Olson~Reichhardt, Phys. Rev. E \textbf{68}, 046102 (2003)

\bibitem{Speer09}
D.~Speer, R.~Eichhorn, P.~Reimann, Phys. Rev. Lett. \textbf{102}, 124101 (2009)

\bibitem{Chacon10}
R.~Chac\'on, A.M. Lacasta, Phys. Rev. E \textbf{82}, 046207 (2010)

\bibitem{Mukhopadhyay18}
A.K. Mukhopadhyay, B.~Liebchen, P.~Schmelcher, Phys. Rev. Lett. \textbf{120},
  218002 (2018)

\bibitem{Reichhardt02a}
C.~Reichhardt, C.J. Olson, Phys. Rev. B \textbf{65}, 174523 (2002)

\bibitem{Tierno07}
P.~Tierno, T.H. Johansen, T.M. Fischer, Phys. Rev. Lett. \textbf{99}, 038303
  (2007)

\bibitem{Soba08}
A.~Soba, P.~Tierno, T.M. Fischer, F.~Sagu\`es, Phys. Rev. E \textbf{77}, 060401
  (2008)

\bibitem{Ao93}
P.~Ao, D.J. Thouless, Phys. Rev. Lett. \textbf{70}, 2158 (1993)

\bibitem{Yabu97}
H.~Yabu, H.~Kuratsuji, Found. Phys. \textbf{27}, 1585 (1997)

\bibitem{Groszek18}
A.J. Groszek, D.M. Paganin, K.~Helmerson, T.P. Simula, Phys. Rev. A
  \textbf{97}, 023617 (2018)

\bibitem{Pribiag07}
V.S. Pribiag, I.N. Krivorotov, G.D. Fuchs, P.M. Braganca, O.~Ozatay, J.C.
  Sankey, D.C. Ralph, R.A. Buhrman, Nature Phys. \textbf{3}, 498 (2007)

\bibitem{Bolte08}
M.~Bolte, G.~Meier, B.~Kr\"uger, A.~Drews, R.~Eiselt, L.~Bocklage, S.~Bohlens,
  T.~Tyliszczak, A.~Vansteenkiste, B.~Van~Waeyenberge et~al., Phys. Rev. Lett.
  \textbf{100}, 176601 (2008)

\bibitem{Wiersig01}
J.~Wiersig, K.H. Ahn, Phys. Rev. Lett. \textbf{87}, 026803 (2001)

\bibitem{Khoury08}
M.~Khoury, A.M. Lacasta, J.M. Sancho, A.H. Romero, K.~Lindenberg, Phys. Rev. B
  \textbf{78}, 155433 (2008)

\bibitem{vanZuiden16}
B.C. van Zuiden, J.~Paulose, W.T.M. Irvine, D.~Bartolo, V.~Vitelli, Proc. Natl.
  Acad. Sci. (USA) \textbf{113}, 12919 (2016)

\bibitem{Han17}
M.~Han, J.~Yan, S.~Granick, E.~Luijten, Proc. Natl. Acad. Sci. (USA)
  \textbf{114}, 7513 (2017)

\bibitem{Reichhardt19a}
C.~Reichhardt, C.J.O. Reichhardt, Phys. Rev. E \textbf{100}, 012604 (2019)

\bibitem{Yazdi02}
S.~Yazdi, J.L. Aragones, J.~Coulter, A.~Alexander-Katz (2020),
  \texttt{arXiv:2002.06477}

\bibitem{Muhlbauer09}
S.~M{\" u}hlbauer, B.~Binz, F.~Jonietz, C.~Pfleiderer, A.~Rosch, A.~Neubauer,
  R.~Georgii, P.~B{\" o}ni, Science \textbf{323}, 915 (2009)

\bibitem{Yu10}
X.Z. Yu, Y.~Onose, N.~Kanazawa, J.H. Park, J.H. Han, Y.~Matsui, N.~Nagaosa,
  Y.~Tokura, Nature (London) \textbf{465}, 901 (2010)

\bibitem{Nagaosa13}
N.~Nagaosa, Y.~Tokura, Nature Nanotechnol. \textbf{8}, 899 (2013)

\bibitem{Schulz12}
T.~Schulz, R.~Ritz, A.~Bauer, M.~Halder, M.~Wagner, C.~Franz, C.~Pfleiderer,
  K.~Everschor, M.~Garst, A.~Rosch, Nature Phys. \textbf{8}, 301 (2012)

\bibitem{Iwasaki13}
J.~Iwasaki, M.~Mochizuki, N.~Nagaosa, Nature Commun. \textbf{4}, 1463 (2013)

\bibitem{Woo16}
S.~Woo, K.~Litzius, B.~Kr{\" u}ger, M.Y. Im, L.~Caretta, K.~Richter, M.~Mann,
  A.~Krone, R.M. Reeve, M.~Weigand et~al., Nature Mater. \textbf{15}, 501
  (2016)

\bibitem{Tekic19}
J.~Teki\'{c}, A.E. Botha, P.~Mali, Y.M. Shukrinov, Phys. Rev. E \textbf{99},
  022206 (2019)

\bibitem{Xiong19}
L.~Xiong, B.~Zheng, M.H. Jin, N.J. Zhou, Phys. Rev. B \textbf{100}, 064426
  (2019)

\bibitem{Reichhardt15a}
C.~Reichhardt, D.~Ray, C.J.O. Reichhardt, Phys. Rev. B \textbf{91}, 104426
  (2015)

\bibitem{Reichhardt15}
C.~Reichhardt, D.~Ray, C.J.O. Reichhardt, Phys. Rev. Lett. \textbf{114}, 217202
  (2015)

\bibitem{Jiang17}
W.~Jiang, X.~Zhang, G.~Yu, W.~Zhang, X.~Wang, M.B. Jungfleisch, J.E. Pearson,
  X.~Cheng, O.~Heinonen, K.L. Wang et~al., Nature Phys. \textbf{13}, 162 (2017)

\bibitem{Litzius17}
K.~Litzius, I.~Lemesh, B.~Kr{\" u}ger, P.~Bassirian, L.~Caretta, K.~Richter,
  F.~B{\" u}ttner, K.~Sato, O.A. Tretiakov, J.~F{\" o}rster et~al., Nature
  Phys. \textbf{13}, 170 (2017)

\bibitem{Legrand17}
W.~Legrand, D.~Maccariello, N.~Reyren, K.~Garcia, C.~Moutafis,
  C.~Moreau-Luchaire, S.~Coffin, K.~Bouzehouane, V.~Cros, A.~Fert, Nano Lett.
  \textbf{17}, 2703 (2017)

\bibitem{Zeissler20}
K.~Zeissler, S.~Finizio, C.~Barton, A.J. Huxtable, J.~Massey, J.~Raabe, A.V.
  Sadovnikov, S.A. Nikitov, R.~Brearton, T.~Hesjedal et~al., Nature Commun.
  \textbf{11}, 428 (2020)

\bibitem{Liu13}
Y.H. Liu, Y.Q. Li, J. Phys.: Condens. Matter \textbf{25}, 076005 (2013)

\bibitem{Muller15}
J.~M\"uller, A.~Rosch, Phys. Rev. B \textbf{91}, 054410 (2015)

\bibitem{Buttner15}
F.~B{\" u}ttner, C.~Moutafis, M.~Schneider, B.~Kr{\" u}ger, C.M. G{\" u}nther,
  J.~Geilhufe, C.v.K. Schmising, J.~Mohanty, B.~Pfau, S.~Schaffert et~al.,
  Nature Phys. \textbf{11}, 225 (2015)

\bibitem{Martinez16}
J.C. Martinez, M.B.A. Jalil, New J. Phys. \textbf{18}, 033008 (2016)

\bibitem{GonzalezGomez19}
L.~Gonz\'alez-G\'omez, J.~Castell-Queralt, N.~Del-Valle, A.~Sanchez, C.~Navau,
  Phys. Rev. B \textbf{100}, 054440 (2019)

\bibitem{Salimath19}
A.~Salimath, A.~Abbout, A.~Brataas, A.~Manchon, Phys. Rev. B \textbf{99},
  104416 (2019)

\bibitem{Muller17}
J.~M{\" u}ller, New J. Phys. \textbf{19}, 025002 (2017)

\bibitem{CastellQueralt19}
J.~Castell-Queralt, L.~Gonzalez-Gomez, N.~Del-Valle, A.~Sanchez, C.~Navau,
  Nanoscale \textbf{11}, 12589 (2019)

\bibitem{Tomasello18}
R.~Tomasello, S.~Komineas, G.~Siracusano, M.~Carpentieri, G.~Finocchio, Phys.
  Rev. B \textbf{98}, 024421 (2018)

\bibitem{Fert17}
A.~Fert, N.~Reyren, V.~Cros, Nature Rev. Mater. \textbf{2}, 17031 (2017)

\bibitem{Tomasello14}
R.~Tomasello, E.~Martinez, R.~Zivieri, L.~Torres, M.~Carpentieri, G.~Finocchio,
  Sci. Rep. \textbf{4}, 6784 (2014)

\bibitem{Prychunenko18}
D.~Prychynenko, M.~Sitte, K.~Litzius, B.~Kr\"uger, G.~Bourianoff, M.~Kl\"aui,
  J.~Sinova, K.~Everschor-Sitte, Phys. Rev. Applied \textbf{9}, 014034 (2018)

\bibitem{Reichhardt15b}
C.~Reichhardt, C.J.O. Reichhardt, Phys. Rev. B \textbf{92}, 224432 (2015)

\bibitem{Reichhardt17}
C.~Reichhardt, C.J.O. Reichhardt, Phys. Rev. B \textbf{95}, 014412 (2017)

\bibitem{Feilhauer19}
J.~Feilhauer, S.~Saha, J.~Tobik, M.~Zelent, L.J. Heyderman, M.~Mruczkiewicz
  (2019), \texttt{arXiv:1910.07388}

\bibitem{Vizarim20}
N.P. Vizarim, C.~Reichhardt, C.J.O. Reichhardt, V.P. A. (2020),
  \texttt{arXiv:2001.08835}

\bibitem{Stosic17}
D.~Stosic, T.B. Ludermir, M.V. Milo\ifmmode \check{s}\else
  \v{s}\fi{}evi\ifmmode~\acute{c}\else \'{c}\fi{}, Phys. Rev. B \textbf{96},
  214403 (2017)

\bibitem{Fernandes18}
I.L. Fernandes, J.~Bouaziz, S.~Bl{\" u}gel, S.~Lounis, Nature Commun.
  \textbf{9}, 4395 (2018)

\bibitem{Toscano19}
D.~Toscano, S.A. Leonel, P.Z. Coura, F.~Sato, J. Mag. Mag. Mater. \textbf{480},
  171 (2019)

\bibitem{Saha19}
S.~Saha, M.~Zelent, S.~Finizio, M.~Mruczkiewicz, S.~Tacchi, A.K. Suszka,
  S.~Wintz, N.S. Bingham, J.~Raabe, M.~Krawczyk et~al. (2019),
  \texttt{arXiv:1910.04515}

\bibitem{Menezes19}
R.M. Menezes, J.F.S. Neto, C.C.d.S. Silva, M.V. Milo\ifmmode \check{s}\else
  \v{s}\fi{}evi\ifmmode~\acute{c}\else \'{c}\fi{}, Phys. Rev. B \textbf{100},
  014431 (2019)

\bibitem{Palermo20}
X.~Palermo, N.~Reyren, S.~Mesoraca, A.V. Samokhvalov, S.~Collin, F.~Godel,
  A.~Sander, K.~Bouzehouane, J.~Santamaria, V.~Cros et~al., Phys. Rev. Applied
  \textbf{13}, 014043 (2020)

\bibitem{Chen19}
W.~Chen, L.~Liu, Y.~Ji, Y.~Zheng, Phys. Rev. B \textbf{99}, 064431 (2019)

\bibitem{Chen20}
W.~Chen, L.~Liu, Y.~Zheng (2020), \texttt{arXiv:2002.08865}

\bibitem{Lin13}
S.Z. Lin, C.~Reichhardt, C.D. Batista, A.~Saxena, Phys. Rev. B \textbf{87},
  214419 (2013)

\bibitem{Brown19}
B.L. Brown, U.C. T\"auber, M.~Pleimling, Phys. Rev. B \textbf{100}, 024410
  (2019)

\bibitem{Mali20}
P.~Mali, A.~\v{S}akota, J.~Teki\'{c}, S.~Rado\v{s}evi\'{c}, M.~Panti\'{c},
  M.~Pavkov-Hrvojevi\'{c}, Phys. Rev. E \textbf{101}, 032203 (2020)

\bibitem{Diaz17}
S.A. D\'{\i}az, C.J.O. Reichhardt, D.P. Arovas, A.~Saxena, C.~Reichhardt, Phys.
  Rev. B \textbf{96}, 085106 (2017)

\bibitem{Sato19}
T.~Sato, W.~Koshibae, A.~Kikkawa, T.~Yokouchi, H.~Oike, Y.~Taguchi, N.~Nagaosa,
  Y.~Tokura, F.~Kagawa, Phys. Rev. B \textbf{100}, 094410 (2019)

\end{thebibliography}

\end{document}